\documentclass[twocolumn,nofootinbib,showpacs,preprintnumbers,amsmath,amssymb,floatfix]{revtex4}
\usepackage{graphicx}
\usepackage{dcolumn}
\usepackage{bm}
\usepackage{color}
\usepackage{url}
\usepackage{amsmath}

\newcommand{\Msun}{M_{\odot}}
\newcommand{\pvec}{\vec{\theta}}

\def\gsim{\mathrel{
\rlap{\raise 0.511ex \hbox{$>$}}{\lower 0.511ex
\hbox{$\sim$}}}}
\def\lsim{\mathrel{
\rlap{\raise 0.511ex \hbox{$<$}}{\lower 0.511ex
\hbox{$\sim$}}}}

\newcommand{\h}{\mathcal{H}}

\begin{document}

\title{Bayesian model selection for testing the no-hair theorem 
with black hole ringdowns}
\author{S. Gossan}
\author{J. Veitch}
\author{B. S. Sathyaprakash}
\affiliation{School of Physics and Astronomy, Cardiff University, 5, The Parade, Cardiff, UK, CF24 3YB}

\date{\today}

\begin{abstract}
In this paper we examine the extent to which black hole quasi-normal 
modes (QNMs) could be used to test the no-hair theorem with future ground- 
and space-based gravitational-wave detectors. We model departures from 
general relativity (GR) by introducing extra parameters which change the 
mode frequencies or decay times from their values in GR. With the aid of
Bayesian model selection, we assess the extent to which the presence of 
such a parameter could be inferred, and its value estimated.  We find that 
it is harder to measure the departure of the mode decay times from their 
GR values than it is with the mode frequencies.  The Einstein Telescope 
(ET, a third generation ground-based detector) could detect departures 
of as little as $8\%$ in the frequency of the dominant QNM mode of a 500\,$\Msun$ 
black hole, out to a maximum range of $\simeq 6$\,Gpc ($z\simeq 0.91)$.
The New Gravitational Observatory (NGO, an ESA space mission to detect 
gravitational waves) can detect departures of $\sim 0.6\%$ in a $10^{8}\,\Msun$ 
black hole to a luminosity distance of $50$\,Gpc ($z \simeq 5.1$), and departures 
of $\sim 10\%$ in a $10^{6}\,\Msun$ black hole to a luminosity distance of 
$\simeq 6$\,Gpc.  In this exploratory work we have made a specific choice of 
source position (overhead), orientation (inclination angle of $\pi/3$) and 
mass ratio of progenitor binary ($m_1/m_2=2$).  A more exhaustive Monte Carlo 
simulation that incorporates progenitor black hole spins and a hierarchical 
model for the growth of massive black holes is needed to evaluate a more 
realistic picture of the possibility of ET and NGO to carry out such tests.
\end{abstract}

\maketitle

\section{Introduction}

Merging compact binaries consisting of neutron stars or black holes are 
promising sources \cite{rates:2010cf}
of gravitational radiation for advanced gravitational wave 
detectors that are currently being built \cite{advLIGO:2007,0264-9381-23-19-S01}. These 
sources are also potential laboratories for testing general relativity (GR) in 
the strong field regime 
\cite{Blanchet:1994,Will:2006,SathyaSchutzLivRev09,Mishra:2010tp,YunesPretorius09}.
The result of such an event (with sufficiently massive progenitors) is a 
highly perturbed black hole (BH) which rapidly returns to its quiescent
state (i.e., a Kerr BH \cite{Kerr:1963}), through the emission of 
gravitational waves (GWs).
The perturbation, and therefore the emitted radiation, can be expanded in 
the natural basis of oscillations, the quasi-normal modes (QNMs) 
\cite{Vishu:1970a} (for a recent review on QNM see, e.g., \cite{Berti:2009kk}). A Kerr 
black hole is characterised only by its mass and angular momentum, 
and so are the complex frequencies of its QNM oscillations 
\cite{RuffiniWheeler:1971,Leaver:1985}, although the relative amplitudes 
of the modes depend on the specific details of the excitation.

Detection of the characteristic ringdown GW signal of a BH would,
therefore, allow a direct test of the no-hair theorem \cite{Dreyer:2004}, 
and hence GR, through the comparison of frequencies and decay times of 
these modes with the predictions of GR for a BH with certain 
mass and spin.  In practice, the detection and discrimination of multiple 
modes is essential, as it is first necessary to infer the mass and spin 
of the black hole before checking for consistency between the modes.
If any of the modes have some parameter dependence, other than mass and spin, 
then the mass and spin obtained from these modes will not be consistent 
with that obtained from the others, and thus the source of emission must 
not be a Kerr black hole.

In Ref.\,\cite{Dreyer:2004} Dreyer \emph{et al} first developed the formalism 
for testing GR with QNM. They also suggested the test of the no-hair theorem 
through the measurement of more than one mode. Berti {\em et al,}
\cite{Berti:2005ys,Berti:2007a} investigated the accuracy of measurement 
of individual mode parameters using a Fisher matrix analysis and estimated 
the resolvability of individual modes in the complete signal as a function 
of signal-to-noise ratio. They conclude that the presence of a second mode
can be inferred as long as the signal-to-noise ratio (SNR) 
is larger than a critical value, under the assumption that the
presence of a ringdown signal has been confirmed and the parameters
of the dominant modes are reliably measured. The critical SNR depends on the mass
ratio of the progenitor binary but an SNR of 20 should suffice if the
mass ratio of the progenitor binary is $q=m_1/m_2 \gsim 2.$ 
Kamaretsos \emph{et al} showed recently 
that using black hole ringdown signals that are emitted after a binary merges,
it might be possible to recover the mass ratio of the progenitor binary from
relative amplitudes of the QNMs \cite{Kamaretsos:2011}. Their signal model,
however, is based on non-spinning black hole binaries. Spins might induce
systematic effects that could make it difficult to reliably measure the
mass ratio of progenitor binary. However, we have to await ringdown models for
spinning black hole binaries to assess what might or might not be possible
to measure.  

Kamaretsos \emph{et al} \cite{Kamaretsos:2011} also proposed
different ways of testing GR using QNMs. In particular, they proposed that
it is sufficient to use any three of the observed frequencies and decay times
for checking the consistency of the ringdown signal with GR. This {\em minimal
set} of parameters could make the test far more effective than trying to
resolve the different mode frequencies and decay times. This is one 
approach we shall follow in this work.

More specifically, in this paper we explore the feasibility of the
proposal in Ref.\,\cite{Kamaretsos:2011} for a limited set of sources
and parameter values. Our study is also limited to sources with mass
ratio 1:2.  We find that for the specific ringdown model that neglects progenitor
spins, it is possible to infer the total mass and spin of 
the black hole and the mass ratio of the progenitor binary from the 
observed ringdown GW, using a phenomenological mapping from these 
to the mode parameters. Having established this, we 
extend our waveform model to include arbitrary deviations from the predicted 
values of the frequencies and decay times, which can be estimated 
either along with the physical parameters of the source or on their 
own, if we assume no relationship between the complex QNM frequencies 
and the mass and angular momentum of the BH.
This gives us two methods of testing GR:
\begin{enumerate}
\item Measure the QNM parameters individually and check for 
consistency between the implied mass and angular momentum from each 
of the modes.
\item Assume that the GR waveform is broadly correct but allow 
deviations in the parameters of a subset of the QNMs away from their 
predicted values and then perform model selection between the GR 
and non-GR models.
\end{enumerate}
We study both of these techniques in the context of future space- and 
ground-based detectors, particularly the Einstein Telescope 
\cite{ETdesign,Sathyaprakash:2011bh} and a rescoped version of the 
Laser Interferometer Space Antenna called New Gravitational Observatory (NGO).

The rest of the paper is organised as follows: In Section \ref{s:model} we
define the waveform model and describe the analysis methods used in testing 
the no-hair theorem.  In Section \ref{s:Visibility} we will discuss the 
expected sensitivities of ET and NGO and the distribution of the signal-to-noise 
ratio of ringdown signals in these detectors. Section \ref{s:no-hair tests} 
describes the simulations of the tests of the no-hair theorem using
two different methods: (1) consistency of the various parameters characterizing
a ringdown signal and (2) Bayesian model selection applied to simulated
ringdown signals buried in Gaussian background.
Section \ref{s:conclusions} contains comparisons of the two methods 
used in testing GR and our conclusions.

\section{Signal model and analysis method}\label{s:model}
In this Section we will discuss the nature of the ringdown waveform
used in this study. The main focus will be to use a superposition of
quasi-normal modes containing not only the dominant mode but also the 
first two sub-dominant ones. Ringdown signals that we study are 
assumed to be emitted by deformed black holes that form from the
merger of compact binaries consisting of {\em non-spinning} components
in quasi-circular orbits.
The latter assumption allows us to use a phenomenological waveform model
based on numerical relativity simulations; the observed
signal is characterised by only three intrinsic parameters: the mass
and spin of the final black hole and the mass ratio of the progenitor
binary.

\subsection{Ringdown model}

A perturbed black hole emits a spectrum of modes characterized by
three numbers $(l,\,m,\,n).$ Indices $l=2,3,\ldots,$ 
and $m=-l,\ldots,+l,$ are the well-known spherical 
harmonic indices, and $n=0,1,2,\ldots,$ is the mode overtone
index. Overtones other than the fundamental $n=0$ mode are not
excited with significant amplitudes and have much shorter damping times
\cite{Berti:2005ys}.  We shall therefore only consider
the fundamental mode and drop the index from further discussion.

The two polarisations of the gravitational waveform, $h_{+} $ and $ h_{\times} $, 
emitted by a BH of mass $M$ during its ringdown are described as the sum over the QNMs,
\begin{align}
 h_{+}(t) = \frac{M}{r}\sum_{l,m > 0} A_{l|m|}\, e^{-t/\tau_{lm}} \,
 Y^{lm}_{+} \cos(\omega_{lm}t - m\phi) \label{h_plus}, \\
h_{\times}(t) = - \frac{M}{r}\sum_{l,m > 0} A_{l|m|}\, e^{-t/\tau_{lm}} \,
Y^{lm}_{\times} \sin(\omega_{lm}t - m\phi),
\label{h_cross}
\end{align}
for $t\ge 0$ and $h_+=h_\times=0,$ for $t<0,$ $t=0$ being the start of the 
ringdown signal. Here $r$ is the luminosity distance to the black hole, $ A_{l|m|}(q)$ are 
the mode amplitudes that depend only on the ratio $q=m_1/m_2$ $(m_1>m_2)$ 
of the component masses of the progenitor binary, and $ \tau_{lm}(M,j)$ 
and $ \omega_{lm}(M,j) $ are the characteristic mode damping times and 
frequencies.  The mode damping times and frequencies depend only
the black hole mass $M$ and its spin $j,$ but they are difficult
to compute analytically. It is necessary to use numerical
methods to compute them \cite{Berti:2009kk}. Fits to some of the lower order modes
can be found in Berti et al \cite{Berti:2005ys}. 
$\iota\in[0,\pi)$ is the angle between the BH 
spin-axis and the line-of-sight to the observer and $\phi\in[0,2\pi]$ is the azimuth
angle of the black hole with respect to the observer. 
$Y^{lm}_{+,\times}(\iota)$ are the sum of spin $-2$ weighted spherical harmonics
\cite{Berti:2007a},
\begin{align}
Y_+^{lm}(\iota)\equiv{}_{-2}Y^{lm}(\iota,0)+(-1)^l~ _{-2}Y^{l-m}(\iota,0),\\
Y_\times^{lm}(\iota)\equiv{}_{-2}Y^{lm}(\iota,0)-(-1)^l~ _{-2}Y^{l-m}(\iota,0).
\end{align}

Although the waveform in Eqs. (\ref{h_plus}) and (\ref{h_cross}) contains 
a summation of modes over all values of $l$ and $m $, the relative 
amplitudes of the higher order modes are significantly less than those 
of the lower, so we restrict the sum to the most significant modes. The 
waveform considered for this analysis is a superposition of the  
$n = 0$, $l = 2$, $m = 1 $ and $n = 0$, $l = m = \{2, 3, 4\}$ 
modes (which will hereafter be referred to as the 21, 22, 33 and 44 
modes, respectively) with the 22 mode being dominant. 

\begin{figure}
\includegraphics[width=0.95\columnwidth]{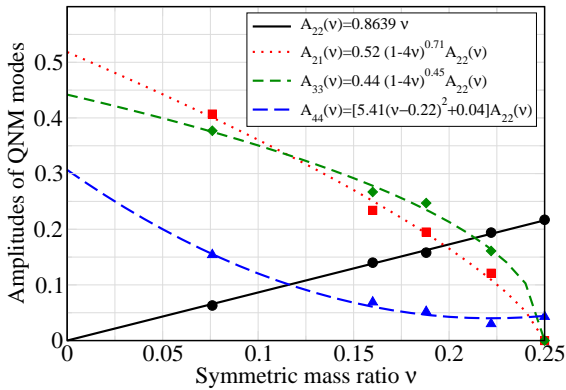}
\caption{Fits to the amplitudes of modes excited in the process of a binary black hole
merger as a function of the symmetric mass ratio of the progenitor binary.}
\label{fig:fits}
\end{figure}

The relative dominance of the higher order modes is dependent on the 
specific excitation that occurred (through $A_{lm}$) and the position 
of the observer (through $Y_{lm}$). This depends on both the mass ratio 
$q=m_1/m_2\ge 1$ of the compact binary components prior to merger, and 
the inclination angle $\iota$. We used a phenomenological model based 
on fits to a set of numerical simulations to describe the mode amplitudes. 
We used the data in Kamaretsos \emph{et al}~\cite{Kamaretsos:2011}, but
instead of using the mass ratio $q$ we used the symmetric mass ratio
$\nu=m_1\,m_2/(m_1+m_2)^2=q/(1+q)^2$ to derive new physically motivated
fits.  The 22 mode is expected to grow linearly with the symmetric mass 
ratio as noted in Refs.\ \cite{Davis:1971,Flanagan:1997sx}.
Modes with odd $l$ or $m$ are not expected to be excited when the 
progenitor binary is of equal mass and so we expect their behaviour
close to $\nu=1/4$ to be some power of $(1-4\,\nu).$ These considerations
led us to the following forms of the various amplitudes:
\begin{eqnarray}
A_{22} (\nu) & = & 0.864\,\nu, \label{amp22} \\
A_{21} (\nu) & = & 0.52\left ( 1 - 4\,\nu \right )^{0.71}  \,A_{22}(\nu),  \label{amp2122} \\
A_{33} (\nu) & = & 0.44\left ( 1 - 4\,\nu \right )^{0.45}  \,A_{22}(\nu),  \label{amp3322} \\
A_{44} (\nu) & = & [5.4 \left (\nu - 0.22 \right )^2 + 0.04] \,A_{22}(\nu).  \label{amp4422} 
\end{eqnarray}
Figure \ref{fig:fits} plots the above fits together with the amplitudes
of the various modes derived from numerical simulations.
The current data seems to indicate that $\hat{A}_{44}\equiv A_{44}(\nu)/A_{22}(\nu)$ has a minimum at $\nu=0.22.$ 
However, the amplitude of the 44 mode derived from numerical simulations is not 
very reliable and so this might be just an artifact of bad data. More accurate
simulations are needed to validate this result.

To complete the signal model we have to specify the damping constants $\tau_{\ell m}.$
Instead of damping times $\tau_{\ell m}$ it is customary to use {\em quality factors}
$ Q_{\ell m}$ defined by $Q_{\ell m} = \omega_{\ell m}\tau_{\ell m}/2$. Berti et al \cite{Berti:2005ys}
provide a simple fitting formula to many mode frequencies and quality factors of the form:
\begin{align}
M\omega &= f_{1} + f_{2}(1 - j)^{f_{3}} \label{m_omega} \\
Q &= q_{1} + q_{2}(1 - j)^{f_{3}} \label{q}
\end{align}
where $ f_{1}$, $f_{2}$, $f_{3}$, $q_{1}$, $q_{2} $ and $ q_{3} $ are 
fitting constants. For the modes considered in this paper Table \ref{fitting_table} lists
the fitting constants.
\begin{table}[h]
\begin{center}
\begin{tabular}{|c|c|c|c|c|c|c|}
\hline
$(l,m)$  & $f_1$ & $f_2$ & $f_3$ & $q_1$ & $q_2$ & $q_3$ \\
\hline
 $ (2,\,1)$  & 0.6000 & $-0.2339$ & 0.4175 & $-0.3000$ & 2.3561 & $-0.2277$ \\
\hline
 $ (2,\,2)$  & 1.5251 & $-1.1568$ & 0.1292 & 0.7000 & 1.4187 & $-0.4990$ \\
\hline
 $ (3,\,3)$  & 1.8956 & $-1.3043$ & 0.1818 & 0.9000 & 2.3430 & $-0.4810$ \\
\hline
 $ (4,\,4)$  & 2.3000 & $-1.5056$ & 0.2244 & 1.1929 & 3.1191 & $-0.4825$ \\
\hline
\end{tabular}
\caption{The fitting constants in Eqns. (\ref{m_omega}) and 
(\ref{q}), for the 21, 22, 33 and 44 modes, for the dominant $n=0$ overtone 
\cite{Berti:2005ys}.}
\label{fitting_table}
\end{center}
\end{table}

The gravitational-wave strain observed by a detector due to the excitation 
of QNM can be expressed as
\begin{equation}
h = F_{+}h_{+} + F_{\times}h_{\times}
\label{h}
\end{equation}
where $F_{+}$ and $F_{\times}$ are the detector antenna pattern functions, 
which depend on the position $(\theta,\varphi)$ of the black hole relative
to the detector and polarisation angle $\psi$ of the radiation 
(see Ref.\, \cite{Kamaretsos:2011} for details). 
The measured strain, therefore, depends on the following 9 parameters:
\begin{equation}
\vec \theta = \left\{
M,\,\nu,\,r,\,\theta,\,\varphi,\,\psi,\,\iota,\,\phi,\, t_0
\right\},
\label{eq:params}
\end{equation}
which includes the epoch $t_0$ when the signal arrives at our
detector\footnote{Note that Eqs.\,(\ref{h_plus}) and (\ref{h_cross})
are written assuming $t_0=0.$ A nonzero value of $t_0$ can be trivally 
included in the signal model by writing $t$ as $t-t_0$ and assuming
$h_{+,\times}(t)=0$ for $t<t_0.$}.

\begin{figure*}
\includegraphics[width=0.48\textwidth]{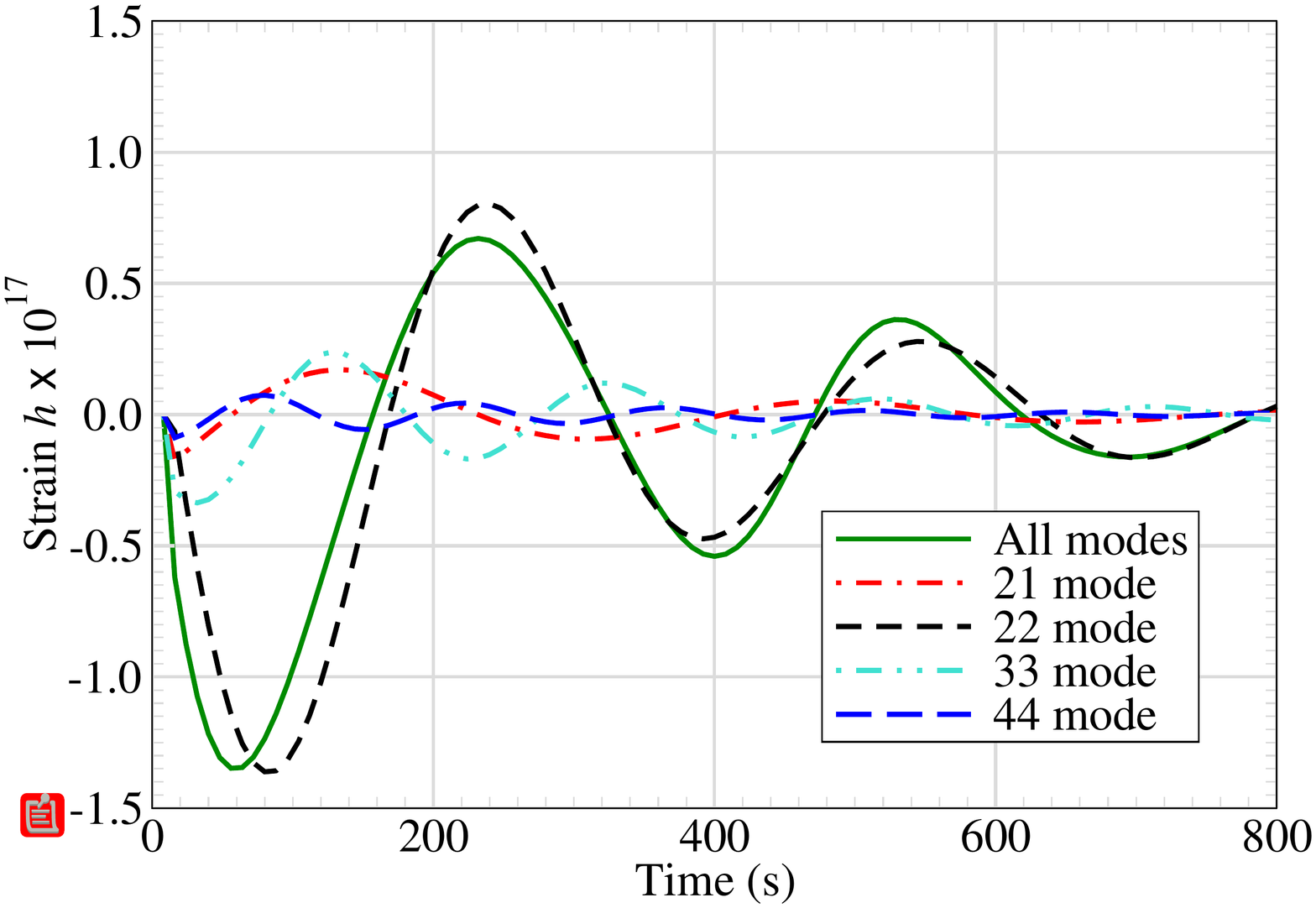}
\includegraphics[width=0.48\textwidth]{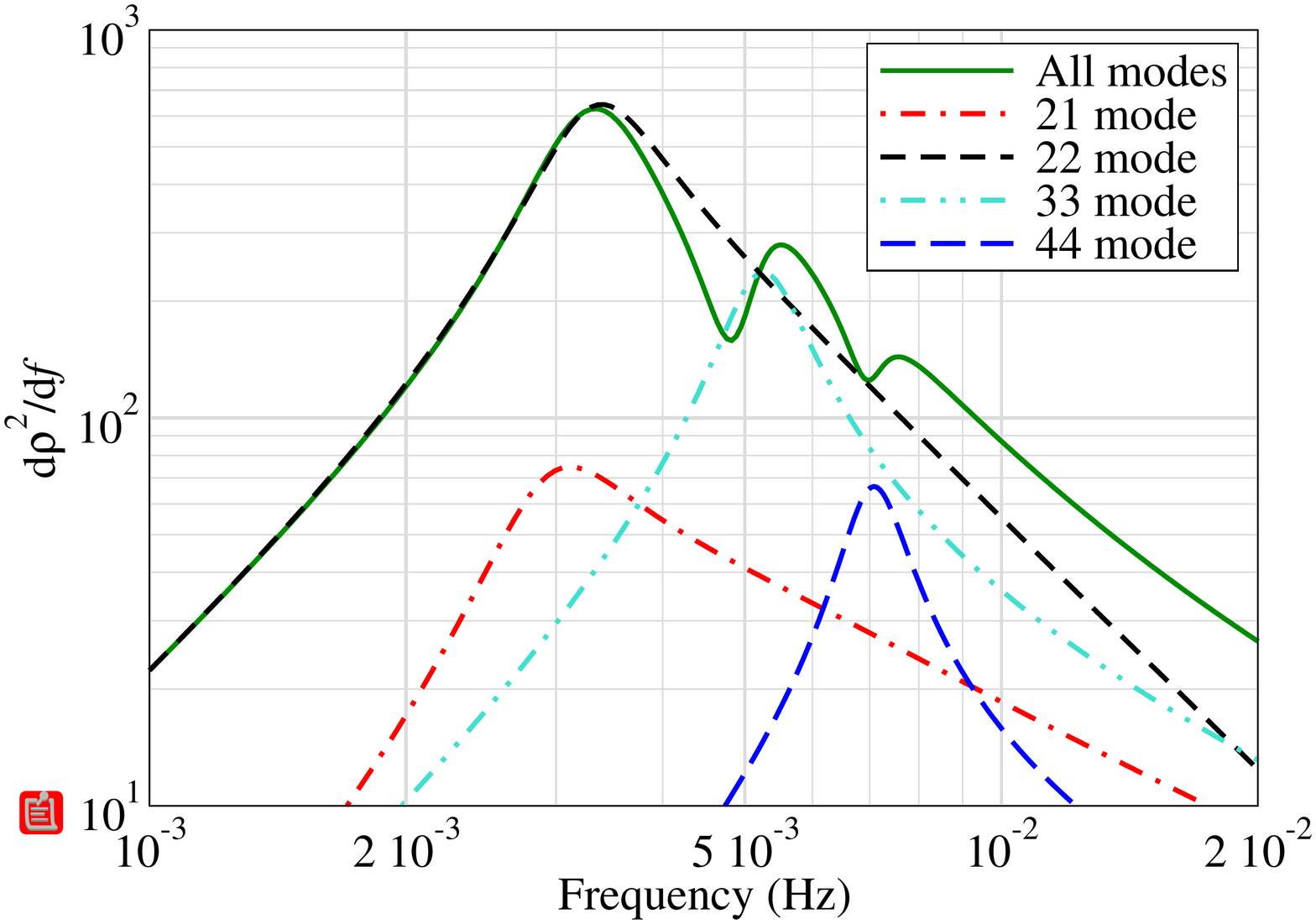}
\caption{{\em Left}: Strain amplitude of a quasi-normal mode signal from a black hole
that forms from the merger of a binary of (observed) total mass 
$5\times 10^6\,M_\odot$ and mass ratio $q=2$ at 1 Gpc. We have plotted the first four
dominant modes, 21, 22, 33 and 44 together with their superposition. 
{\rm Right}: The signal-to-noise ratio
integrand of the same signal $d\rho^2/df=|H(f)|^2/S_h(f),$ where $S_h(f)$ is
taken to be that of NGO. The presence of the 33 and 44 mode can be clearly
seen in the overall spectrum, 21, however, is buried under 22.}
\label{fig:strains and spectra}
\end{figure*}

Figure \ref{fig:strains and spectra}, left panel, shows the response of
a signal containing all four modes considered in this work as well as
the phase evolution of each of modes 21, 22, 33 and 44 for a $5\times 10^6
M_\odot$ black hole at a distance of 1 Gpc, assumed to be formed from a
binary of mass ratio $q=2.$ The black hole is assumed to be overhead 
with respect to the detector, the orbital plane
making an angle of $\iota=\pi/3$ radians --- a sub-optimal orientation but 
one for which the sub-domimant modes will have a nonzero
amplitude. The polarization and azimuth angles are randomly
chosen to be $\psi=2.67$ radians and $\phi=2.31$ radians. 

The overall amplitude is very nearly the same as that of the 22 mode.
The amplitude of the 33 mode, which is the second most dominant 
after 22, is about a third
of the 22 mode, followed by the 21 mode, which has a slightly smaller
amplitude. The 44 mode, whose amplitude is about 12\% of the 22 mode, has
negligible effect on the overlall signal.
Although 21 and 33 are of roughly the same amplitude, the energy in 33 
mode, whose frequency is roughly twice that of 21, will be significantly 
larger  as we will see in Sec.\, \ref{s:Visibility}, while discussing the relative
signal-to-noise ratios of the different modes. The interference
between the different modes causes features that become apparent in the
spectrum of the modes shown in the left hand panel of Figure 
\ref{fig:strains and spectra}, which we will discuss further in Sec.\, \ref{s:Visibility}.

\subsection{A generalised model for the QNM}

In order to test GR, we extended the waveform model so that
the frequencies and decay times of the modes were allowed to be 
dependent not only on $M$ and $j$, but also other 
dimensionless parameters. More specifically, we considered that
frequencies $\omega_{lm}$ depended on three parameters 
$(M,\,j,\,\Delta\hat{\omega}_{lm})$ and damping times $\tau_{lm}$
also depended on three parameters $(M,\,j,\,\Delta\hat{\tau}_{lm}).$ 
Furthermore, we assumed that the dimensionless parameters
$\Delta\hat{\omega}_{lm}$ and $\Delta\hat{\tau_{lm}}$
were independent for each mode. Following Li et al \cite{Li:2011cg},
in this generalised model, the 
frequencies $\omega_{lm,\mathrm{nonGR}}$ and decay times 
$\tau_{lm,\mathrm{nonGR}}$ were expressed as
\begin{eqnarray}
\omega_{lm,\mathrm{nonGR}} & = & \omega_{lm,\mathrm{GR}}
\left (1 + \Delta\hat{\omega}_{lm} \right ) 
\label{deltaomega} \\
\tau_{lm,\mathrm{nonGR}} & = & \tau_{lm,\mathrm{GR}}
\left (1 + \Delta\hat{\tau}_{lm} \right ) 
\label{deltatau}
\end{eqnarray}
where $ \omega_{lm,\mathrm{GR}} $ and $ \tau_{lm,\mathrm{GR}} $ 
are the frequencies and decay times of modes as in GR. 
The signal produced by the GR hypothesis is a special case of 
the generalised model, in which 
$ \Delta\hat{\omega}_{lm} = \Delta\hat{\tau}_{lm} = 0 $ for all $ l, m $.

\subsection{Bayesian analysis}
Having described the waveform model and its parameters (contained in 
a parameter vector $\pvec$\,), we will now describe how these parameters 
are estimated from data containing a ringdown signal. We assume that 
the data from the gravitational wave detector in the frequency 
domain $\tilde{d}$ contains both the ringdown signal $\tilde{h}(f;\pvec\,)$ 
and some additive Gaussian noise with known power spectrum $S_h(f)$. Thus,
the data $\tilde{d}$ is assumed to be $\tilde{d}_i=\tilde{h}(f_i;\pvec)+
\tilde{n}_i$, where $i$ is the index of the frequency bin. The noise 
power spectra used for ET and NGO are given in Section 
\ref{s:sensitivity}.
As we perform our analysis in the frequency domain, we use the 
Fourier transformed signal model $\tilde{h}(f) = \int_0^{\infty} 
h(t)e^{-2\pi ift} dt$ computed with the FFTW package.

Our goal is to compute the posterior probability 
distribution (PDF) (see, for instance, Ref.\,\cite{Veitch:2010}), 
of the parameters $p(\pvec|d,\h)$,
\begin{equation}
p(\pvec|d,\h)=\frac{p(d|\pvec,\h)p(\pvec|\h)}{p(d|\h)},
\end{equation}
where $p(d|\h)$ is the evidence, or marginal likelihood, of the model 
\begin{equation}
p(d|\h)=\int_\Theta p(\pvec|\h)p(d|\pvec,\h)d\pvec,
\end{equation}
$\h$, $p(\pvec|\h)$ is the prior distribution of the parameters 
given the signal model and $p(d|\pvec,\h)$ is the likelihood of 
the data for a particular set of parameters $\pvec:$ 
\begin{equation}
p(d|\pvec,\h)\propto\exp\left(-2\sum\frac{|\tilde d_i-\tilde 
h(f_i;\pvec)|^2}{S_h(f_i)}\right).
\end{equation}
Posterior distributions for particular parameters of 
interest, e.g. the $\Delta\hat{\omega}_{lm}$ and $\Delta\hat{\tau}_{lm}$, 
are computed by marginalising the PDF over all other parameters. We also compute the Bayes factor $B_{i,j}$ between
various hypotheses, $B_{i,j}=p(d|i)/p(d|j)$, which is the evidence ratio between hypotheses $i$ and $j$. Due to the large
range of this quantity we will always use the natural logarithm of this, $\log{B_{i,j}}$, in section \ref{models}.

To compute the PDFs and evidence we use the \texttt{inspnest} implementation 
of the nested sampling algorithm, described in \cite{Veitch:2010}, 
modified to use the ringdown signal model, which is available as part of the LSC 
Algorithm Library \cite{LAL}.  The end product of the analysis is the model 
evidence and a set of samples from the posterior PDF that are histogrammed 
to produce the figures below.


\section{Visibility of the QNM Signals}\label{s:Visibility}
In this Section we will explore the visibility of quasi normal
mode signals. We will begin with the expected sensitivities
of ET and NGO and then go on to explore the visibility of the signal
in these two detectors. The signal-to-noise ratio (SNR) of a ringdown
signal depends quite critically on a number of intrinsic and extrinsic 
parameters of the source. We will vary most of these parameters to 
obtain a distribution of the SNR.

\begin{figure*}
\includegraphics[width=0.95\columnwidth]{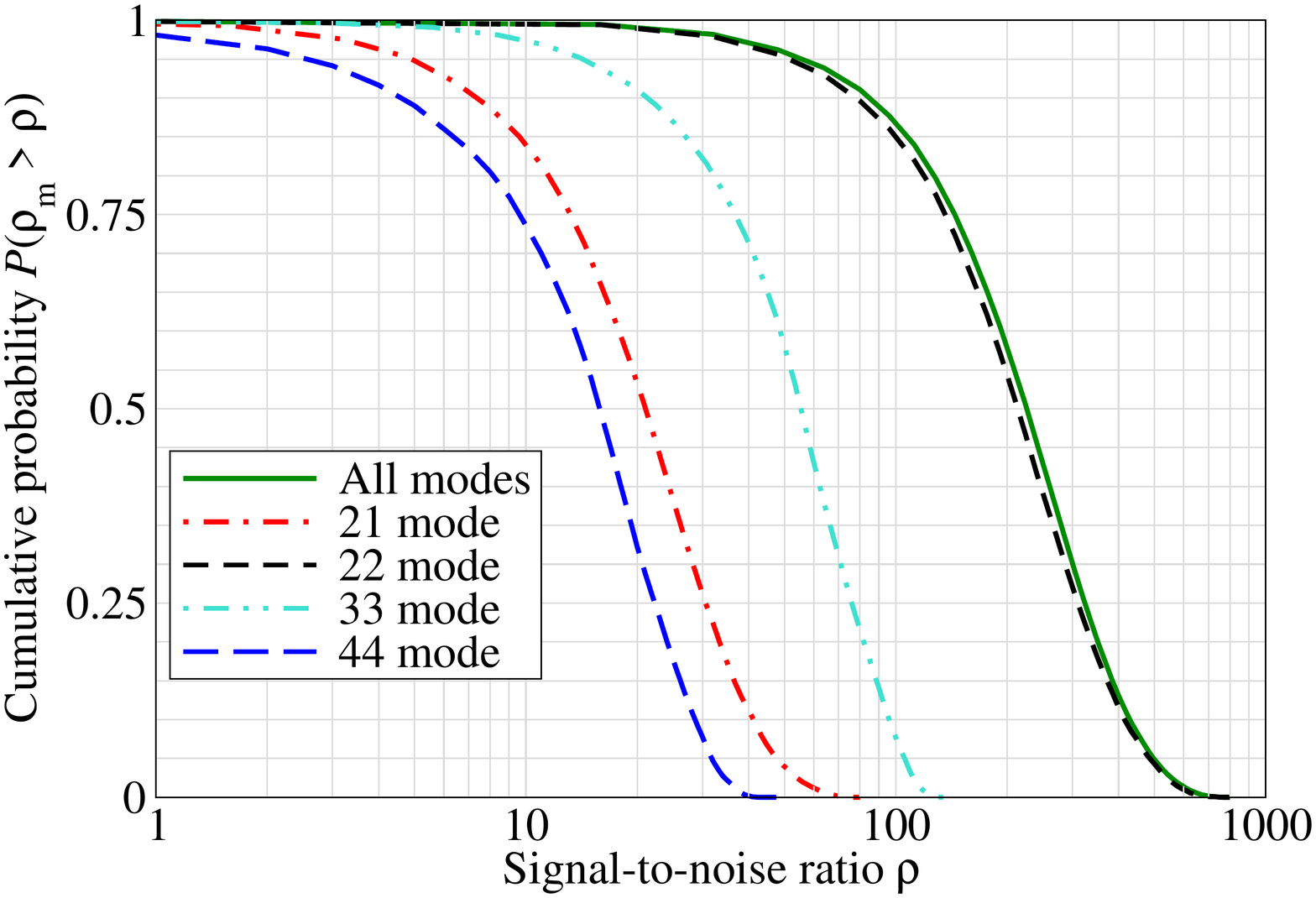}
\includegraphics[width=0.95\columnwidth]{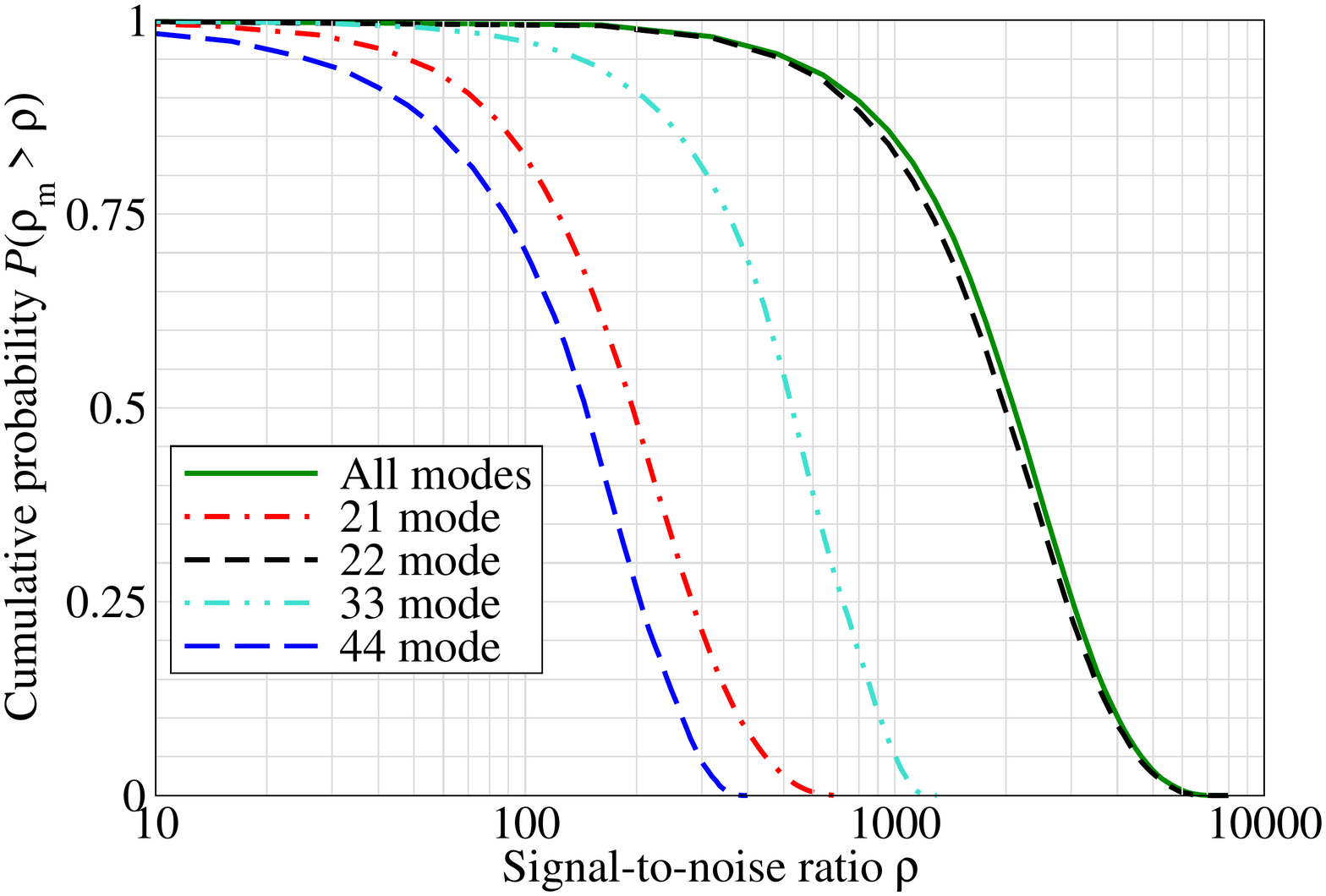}
\caption{Distribution of the signal-to-noise ratios in different quasi-normal
modes (21, 22, 33 and 44) for sources
located at random positions on the sky, with random inclination and
polarisation angles. The left plot is for QNM resutling from the merger
of a 500 solar mass binary observed in ET and the right plot is for
a 5 million solar mass binary observed in NGO. In both cases the
mass ratio of the binary is assumed to be $q=2$ and the source
is assumed to be at 1 Gpc.}
\label{fig:SNR}
\end{figure*}

\begin{figure*}[h]
\centering
$\begin{array}{cc}
\includegraphics[width=\columnwidth]{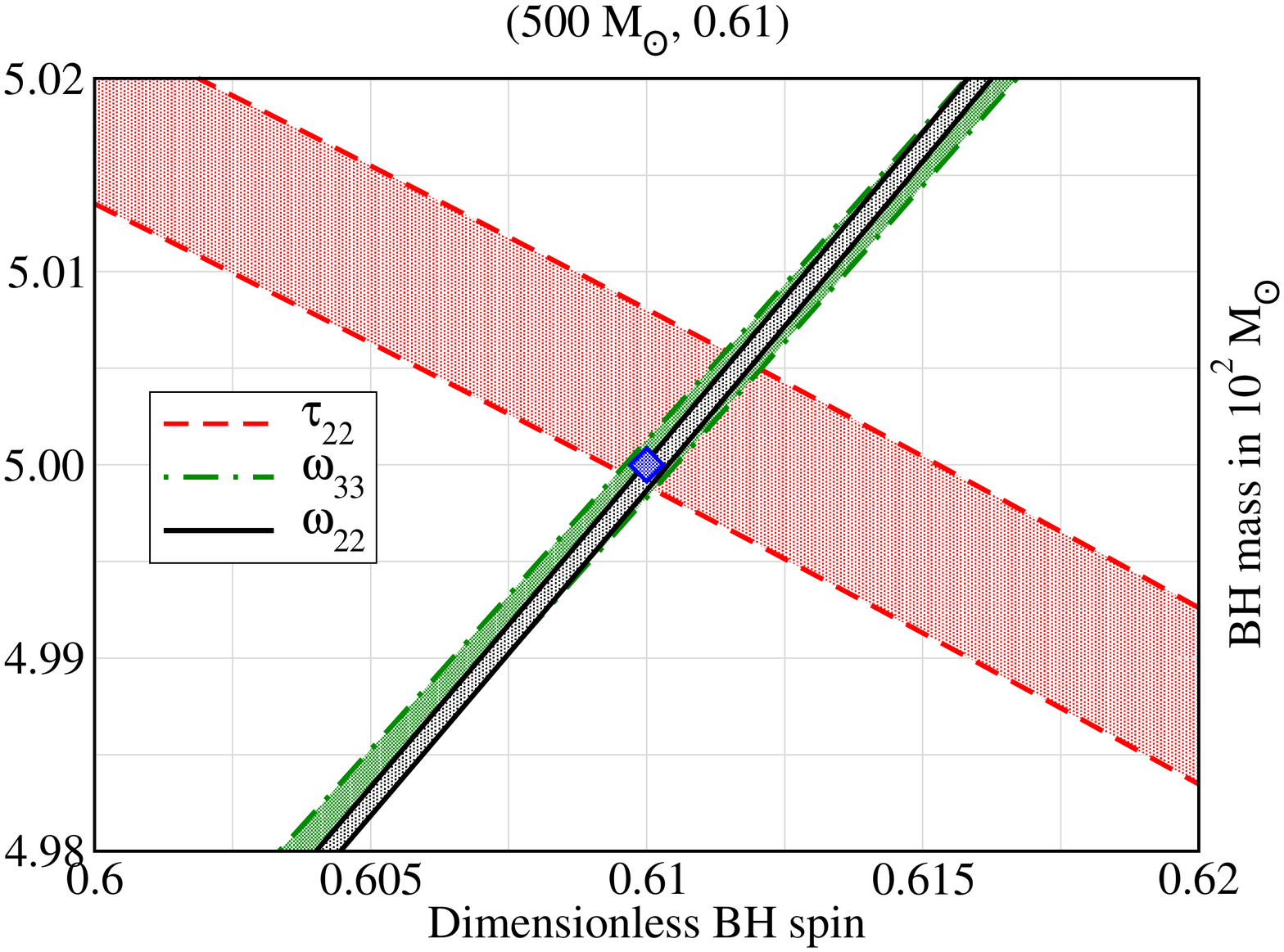} & \includegraphics[width=\columnwidth]{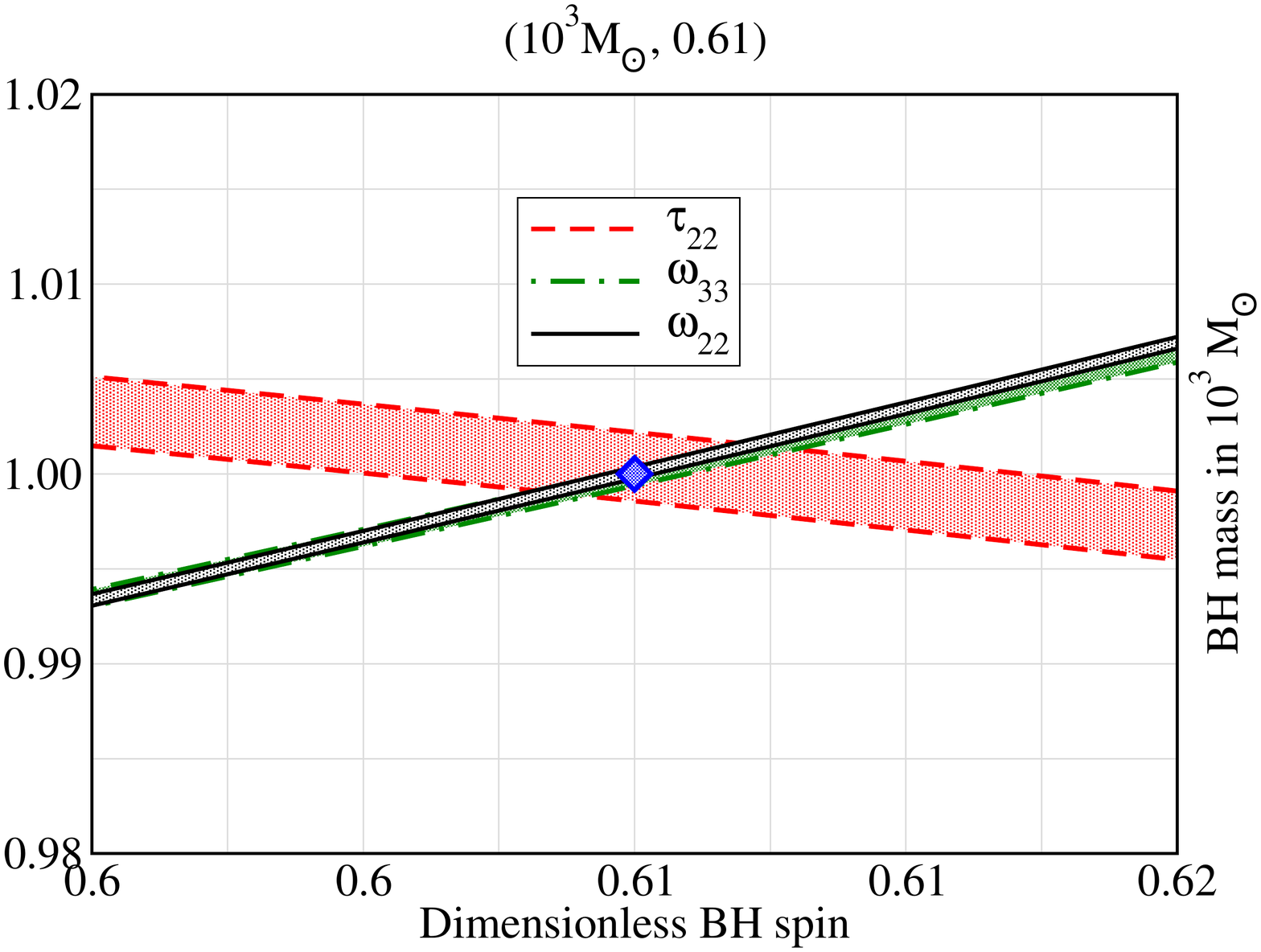} \\
\includegraphics[width=\columnwidth]{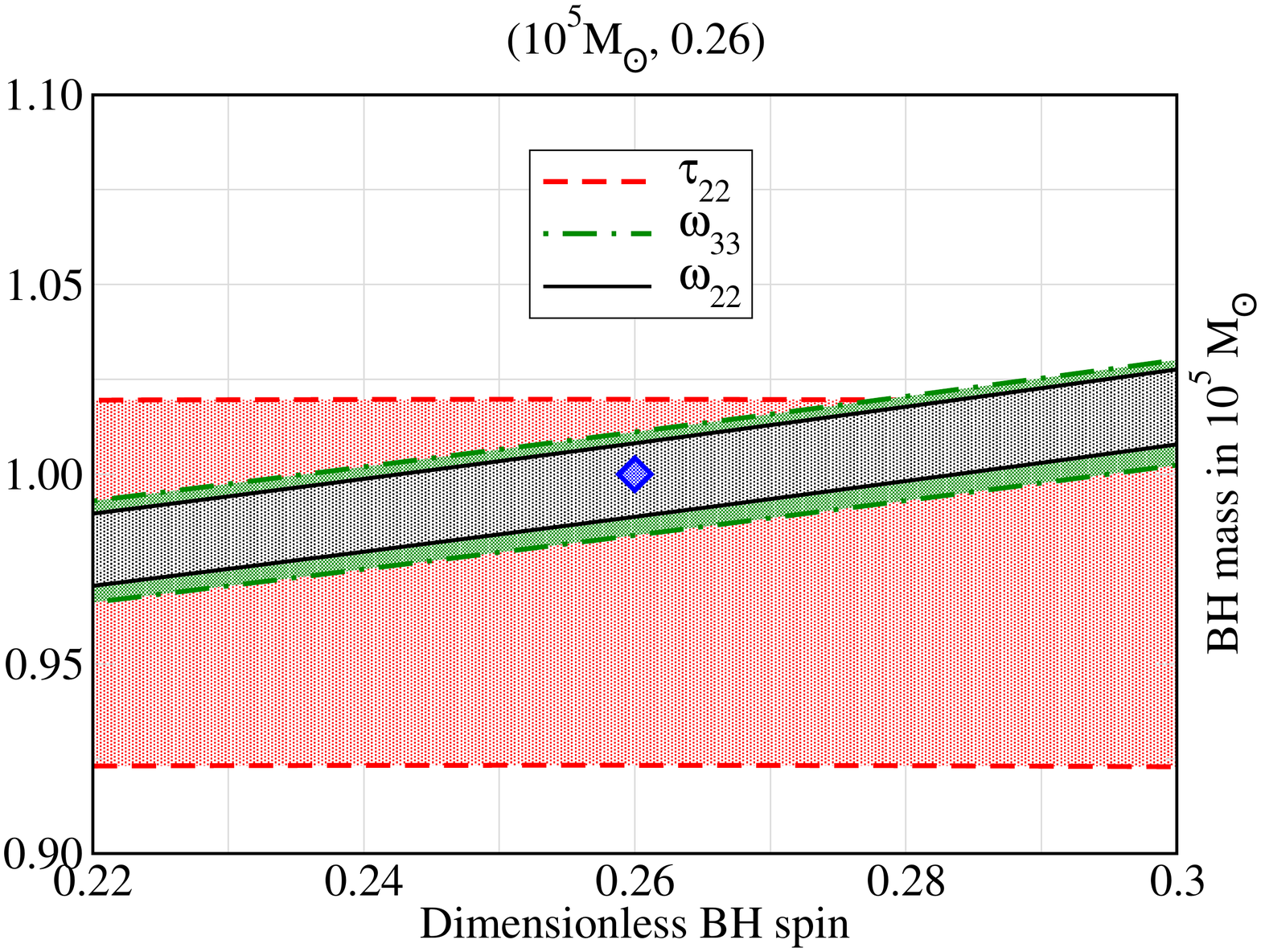} & \includegraphics[width=\columnwidth]{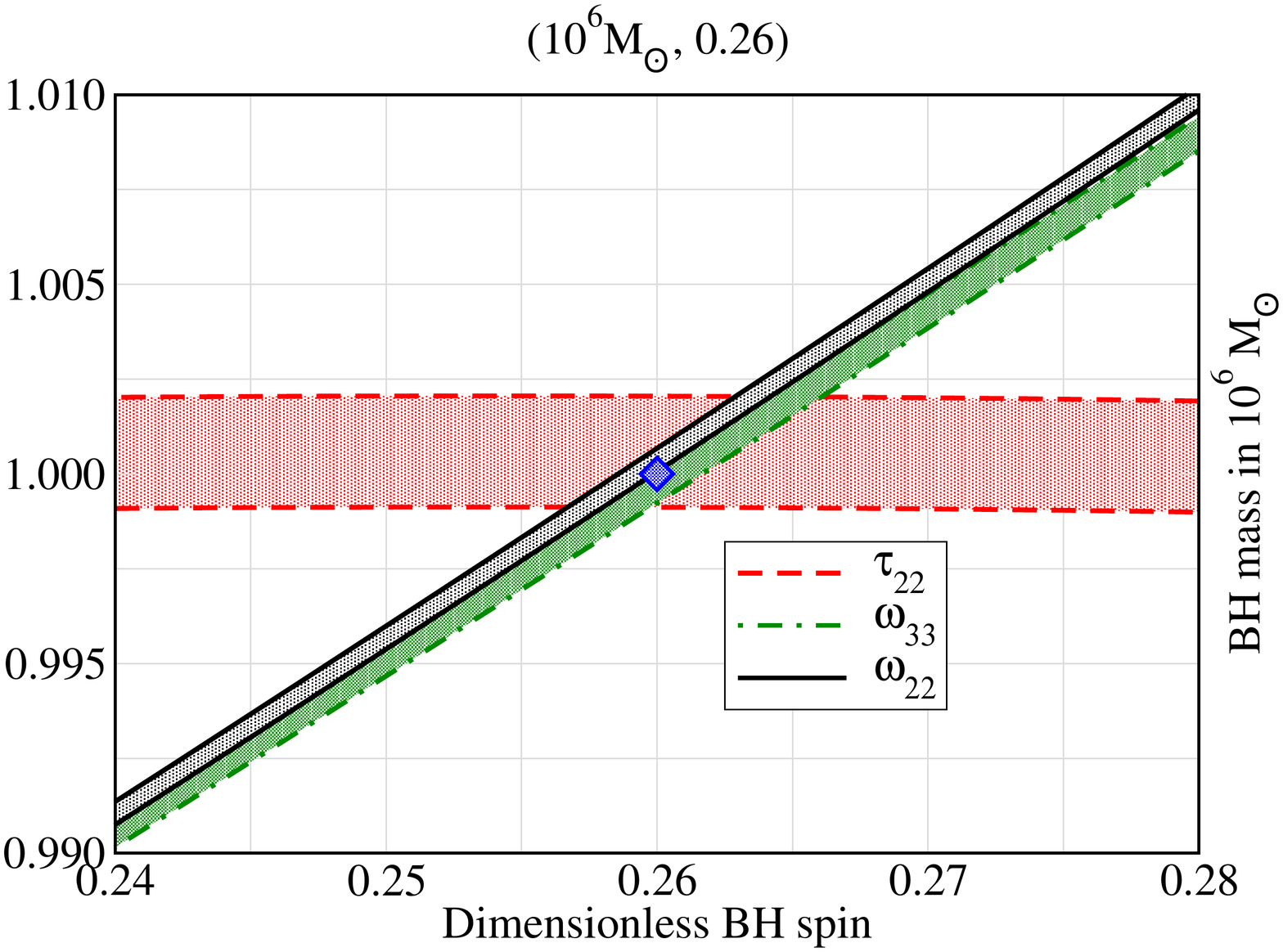} \\
\includegraphics[width=\columnwidth]{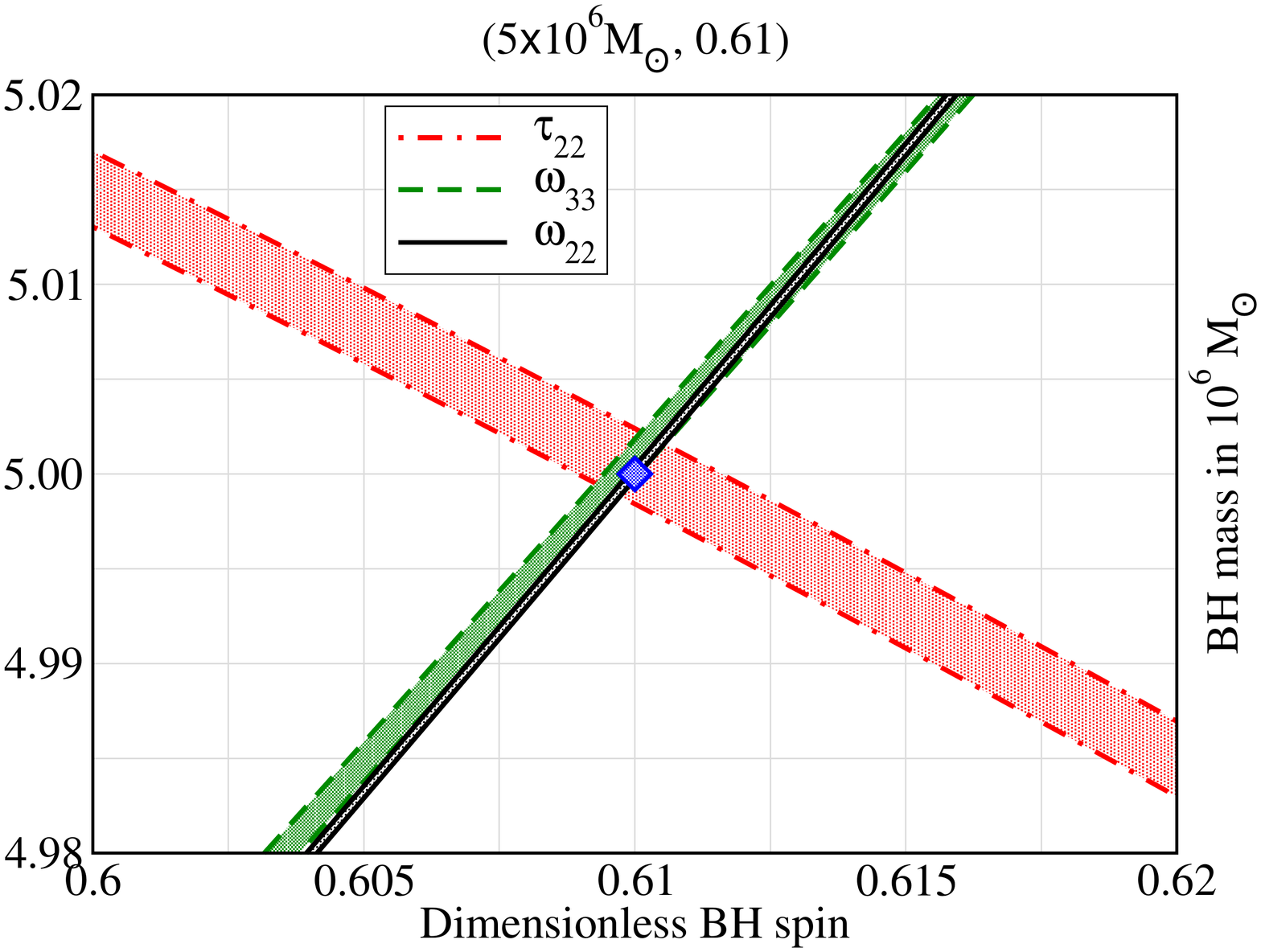} & \includegraphics[width=\columnwidth]{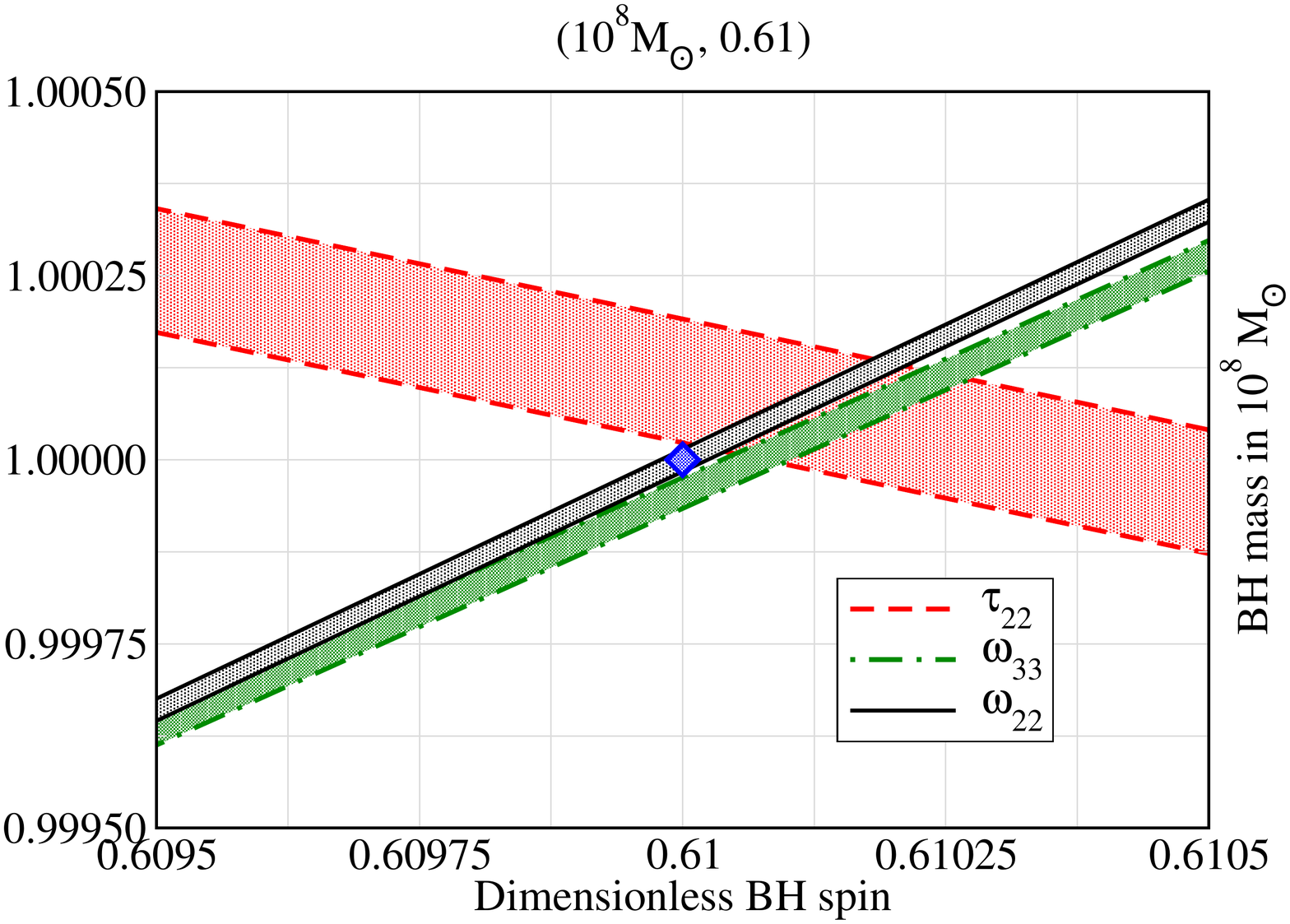}
\end{array}$
\caption{Projections in the $(M, j)$-plane of the 90$\%$ confidence limits on $\omega_{22}$, $\tau_{22}$ and $\omega_{33}$ (blue, blue dotted and red lines respectively)
for injections of signals consistent with GR for $M = 500\,\Msun$ (top-left at 125\,Mpc; SNR = $2\,888$), $M=1000\,\Msun$, (top-right at 225\,Mpc; SNR = $2\,423$); $M=10^{5}\,\Msun$ (middle-left at 125\,Mpc; SNR = $63$), $M=10^{6}\,\Msun$ (middle-right at 125\,Mpc; SNR = $1\,756$), $M=5 \times 10^{6}\,\Msun$ (bottom-left at 1\,Gpc; SNR = $6\,377$) and $M=10^{8}\,\Msun$ (bottom-right at 1\,Gpc; SNR = $115\,154$).  The injected value is denoted in each case by a diamond.}
\label{minset_GRconfidence}
\end{figure*}

\begin{figure*}[h]
\centering
$\begin{array}{cc}
\includegraphics[width=\columnwidth]{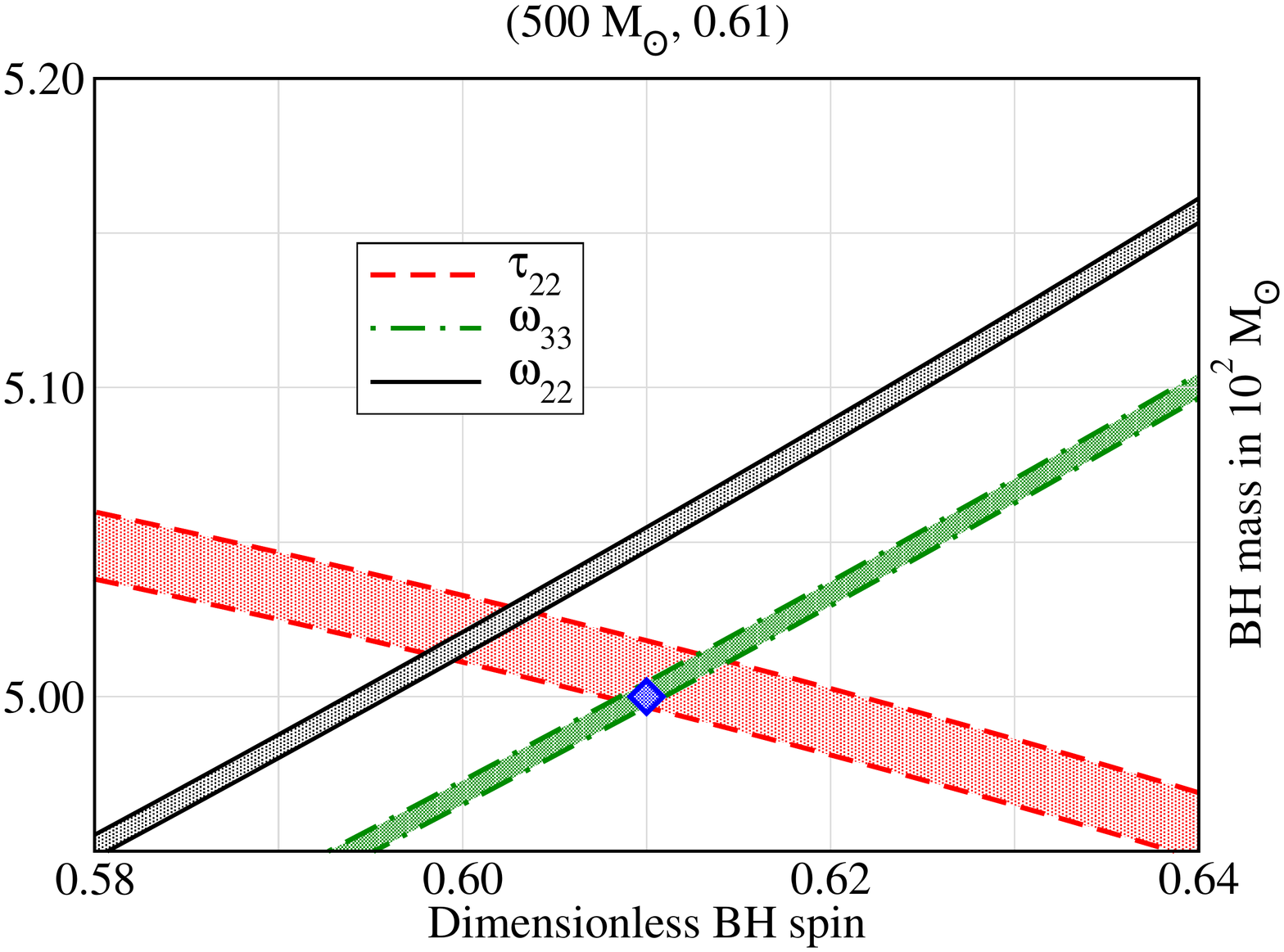} & \includegraphics[width=\columnwidth]{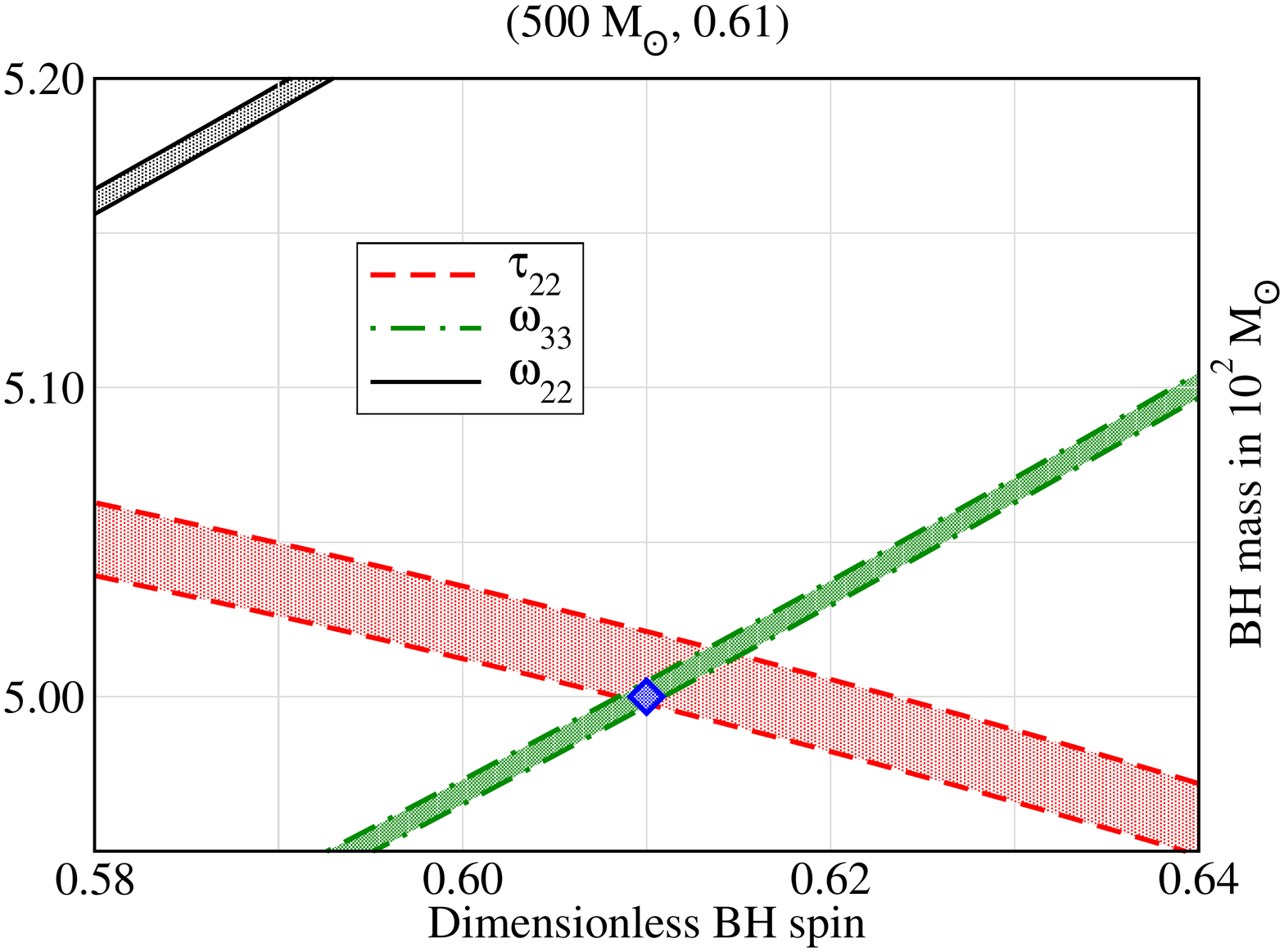} \\
\includegraphics[width=\columnwidth]{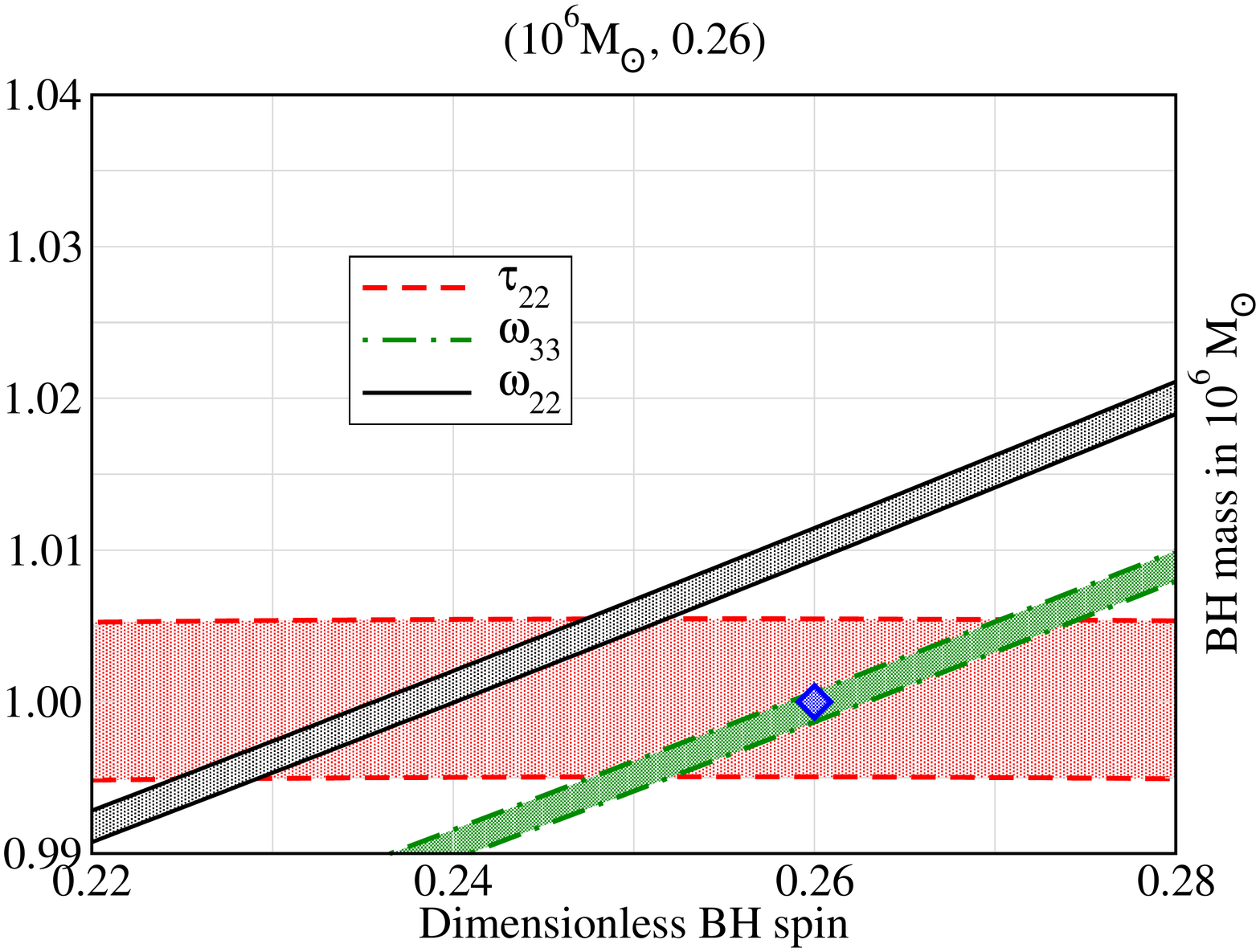} & \includegraphics[width=\columnwidth]{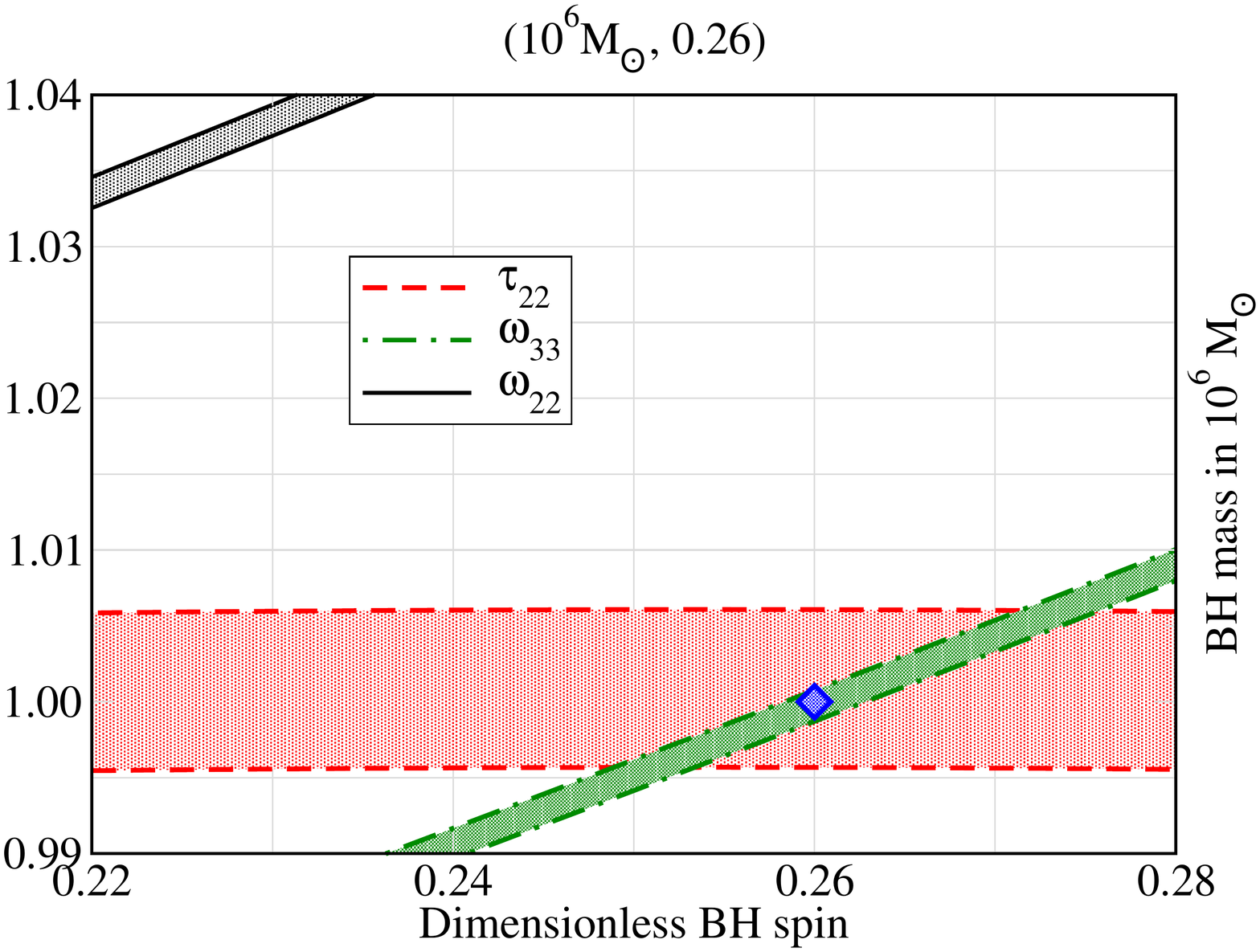} \\
\includegraphics[width=\columnwidth]{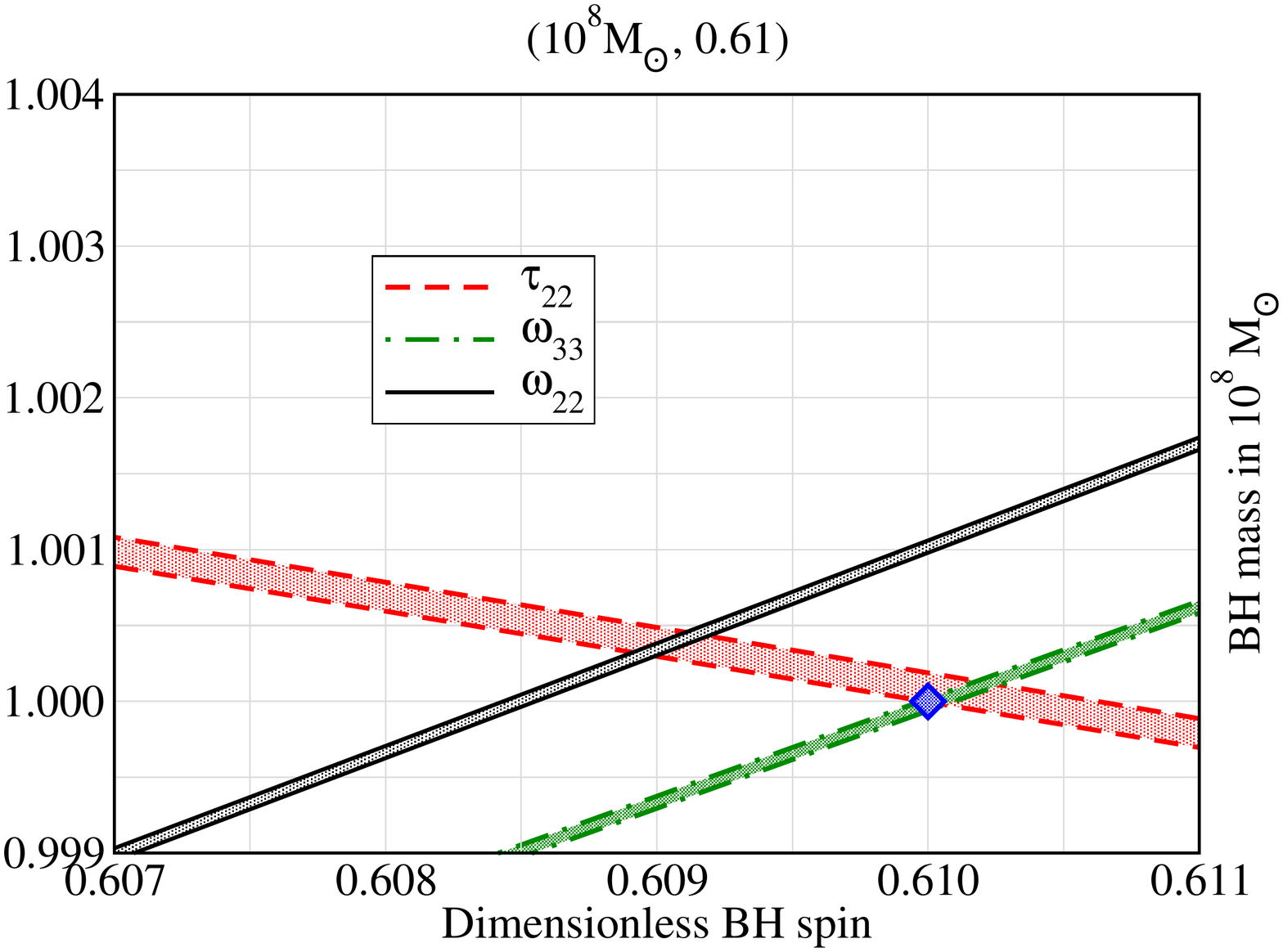} & \includegraphics[width=\columnwidth]{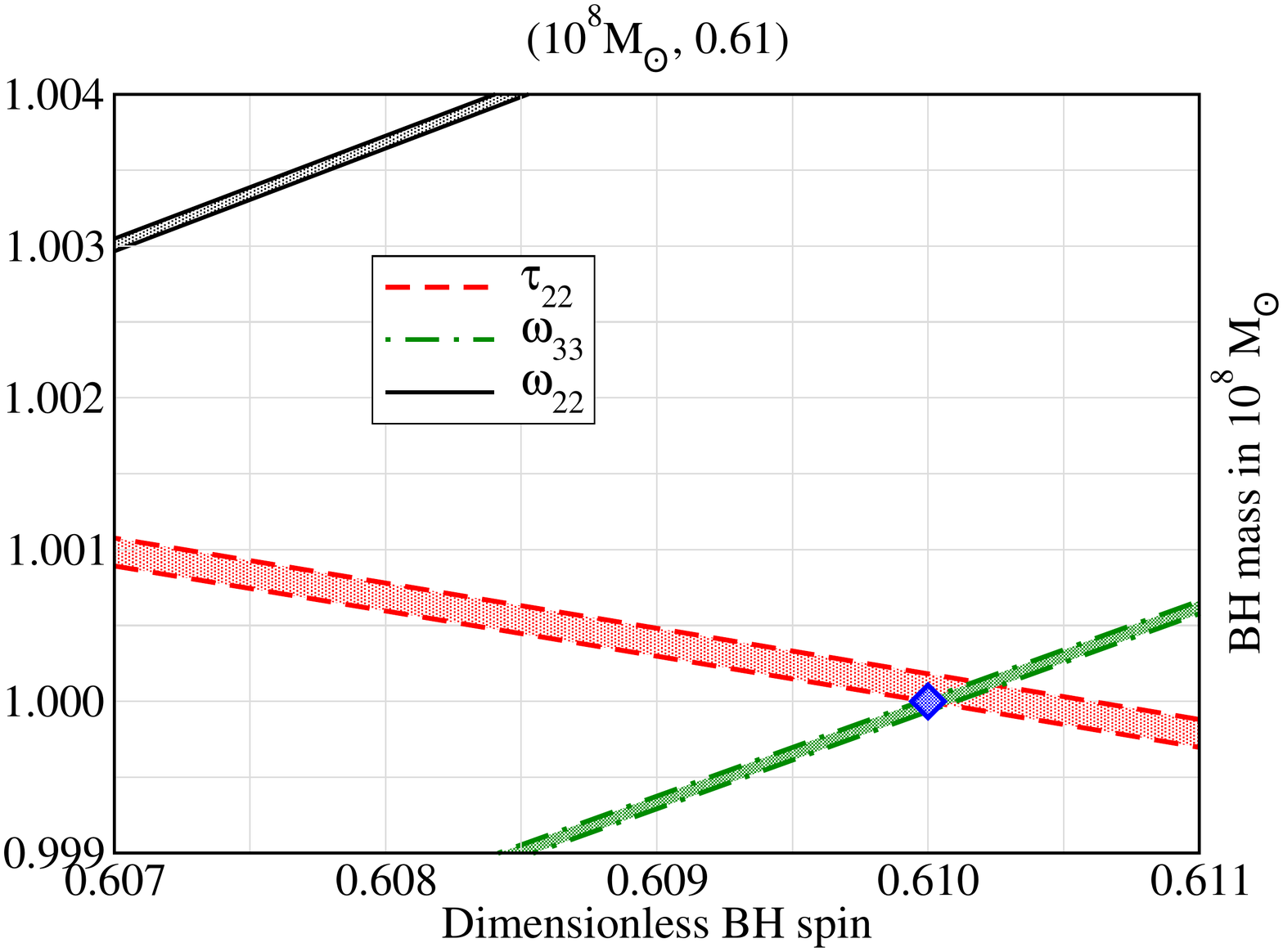} \\
\end{array}$
\caption{Projections in the $(M, j)$-plane of the 90\% confidence limits on $\omega_{22}$, $\tau_{22}$ and $\omega_{33}$ (blue, blue dotted and red lines, respectively)
for non-GR injections of $M = 500\,\Msun$ (top at 125\,Mpc; with $\Delta\hat{\omega}_{22} = -0.01$, SNR = $2\,867$ (left) and $\Delta\hat{\omega}_{22} = -0.05$, SNR = $2\,779$ (right)), $M = 10^{6}\,\Msun$ (middle at 125\,Mpc; with $\Delta\hat{\omega}_{22} = -0.01$, SNR = $1\,753$ (left) and $\Delta\hat{\omega}_{22} = -0.05$, SNR = $1\,735$ (right)) and $M = 10^{8}\,\Msun$ (bottom at 1\,Gpc; with $\Delta\hat{\omega}_{22} = -0.001$, SNR = $115\,130$ (left) and $\Delta\hat{\omega}_{22} = -0.005$, SNR = $115\,031$ (right)).  The injected value is denoted in each case by a diamond.}
\label{minset_nonGRconsistency}
\end{figure*}

\subsection{Sensitivity curves}\label{s:sensitivity}
In this paper, the ET and NGO detectors are considered.  For 
simulations concerning ET, the power spectral density corresponding 
to the ET-B sensitivity curve is considered,
described by $S_{h}(f) = 10^{-50}h_{n}(f)^{2}\,\mathrm{Hz}^{-1}$, 
where $h_{n}(f)$ is given by
\begin{eqnarray*}
h_{n}(f) &=& 2.39\times10^{-27}x^{-15.64} + 0.349x^{-2.145} \\
&& {} + 1.76x^{-0.12} + 0.409x^{1.10},
\label{ETpsd}
\end{eqnarray*}
and $x = f/100$Hz.  For NGO, the sensitivity curve associated with 
the $L=1\times10^9$\,m arm, 4-link mission studied in Ref.\,\cite{NGOref}, 
which corresponds to a power spectral density  given by
\begin{align}
S_{h}(f)=\frac{10}{3L^2}\left(1+ \left[\frac{2Lf}{0.41c}\right]^2\right)\left(4S_{acc} + S_0 \right ), 
\end{align}
where, $S_0=1.153\times10^{-22}\,\mathrm{m^2Hz^{-1}}$ and 
\begin{align}
S_{acc}=1.37\times 10^{-32}\left(1+\frac{10^{-4}\,\mathrm{Hz}}{f}\right)\left(\frac{2\pi f}{1\,\mathrm{Hz}}\right)^{-4}\,\mathrm{m^2Hz^{-1}}.
\label{newLISApsd}
\end{align} 

\subsection{Visibility of ringdown signals}
An important requirement for tests of the no-hair theorem is good visibility of
the signal in our detector: larger the signal-to-noise ratio (SNR), greater
would be the power of the test.  The SNR depends both on the intrinsic parameters of the
black hole, its mass and spin angular momentum, as well as its extrinsic parameters, 
its distance and various angles describing the orientation of its spin axis, its 
location on the sky and the polarisation of the radiation.  In this Section we will 
look at the distribution of the SNR in ET and eLISA/NGO. 

The SNR of a ringdown signal observed using matched filtering is given by
\begin{equation}
\rho^2 = 4 \int_{f_{\rm low}}^{f_{\rm high}} \frac{|H(f)|^2}{S_h(f)} {\rm d}f,
\end{equation}
where $S_h(f)$ is the detector noise power spectral density and $H(f)$ is the 
Fourier transform of the signal in Eq.\,(\ref{h}). The limits on the integrand
can, in principle, be from $0$ to $\infty.$ However, in order to prevent the
``junk" radiation that occurs in the Fourier transform due to abrupt
cutoff of the time-domain signal from corrupting the SNR we choose finite
values for both them. More concretely, we choose $f_{\rm low}$ to be half of the 
smallest signal frequency (which in our case is that corresponding to the 21 mode) and 
$f_{\rm high}$ to be twice the highest signal frequency (which in our case is that
corresponding to the 44 mode). 

$\rho$ is the SNR of the signal containing all the modes considered in this paper.
To test the no-hair theorem it is essential that the SNRs of the 
sub-dominant modes (21, 33, 44) are comparable to the dominant 22 mode. To assess
their importance, we shall also separately consider the SNR $\rho_{\ell m}$
of each mode:
\begin{equation}
\rho_{\ell m}^2 = 4 \int_{f_{\rm low}}^{f_{\rm high}} \frac{|H_{\ell m}(f)|^2}{S_h(f)} {\rm d}f,
\end{equation}
where $H_{\ell m}$ is the Fourier transform of the signal in Eq.\,(\ref{h})
but with $h_+$ and $h_\times$ containingly only the relevant mode.
We will compoare the distribution of $\rho_{\ell m}$ for $\ell m=21, 22, 33 $
and 44 modes but it should be kept in mind that $\rho^2$ is not the sum over
different $\rho^2_{\ell m}.$ Even so, the relative strengths of different
modes gives us an indication of how good we can expect our tests of the 
no-hair theorem to be.

Figure \ref{fig:strains and spectra}, right panel, plots the SNR integrand
$d\rho^2/df$ of the various modes. The overall signal gets most of its
contribution from the 22 mode but other modes significantly alter the
phasing of the signal (left panel) and its spectrum. The overall signal
has bumps caused by the 33 and 44 modes but the 21 mode, whose frequency
is close to that of 22, has little effect on the overall amplitude or the
phasing of the waves. Note that the spectrum of 33 is far larger than 
the 21 mode although they are of similar amplitude. 

Figure \ref{fig:strains and spectra} gives a qualitative understanding that
the subdominant modes have significant effect on the phasing and spectrum
of the signal. But we need to evaluate the distribution of the SNR to
better gauge the relative importance of the different modes.
The response of our detectors to a ringdown signal depends on a total of nine
parameters as enumerated in Eq.\,(\ref{eq:params}). To compute the distribution
of the SNR, however, we don't need to vary all nine parameters. We know that the
SNR depends inversely on the luminosity distance $r$ of the source and so this
can be fixed to any value we wish and we take it to be $r=1\,\rm Gpc.$
The distribution does not depend on the epoch of the signal $t_0$ nor on
the phase $\phi.$ We should study the distribution as a function of $M$ and
$\nu$ while varying the remaining four parameters. However, since our goal 
is to explore the test of the no-hair theorem for a small sample of signals 
with high SNR we will restrict ourselves to only a couple of sources.
We will consider a BH that results from the merger of a 500 $M_\odot$ 
binary in the case of ET and a $5\times 10^6\,M_\odot$ binary in the case 
of NGO. In both cases we take the symmetric mass ratio to be $q=2$ or $\nu=2/9.$

The cumulative distribution of the SNR is shown in
Figure \ref{fig:SNR}. We see that there is a 25\% chance that the
SNR will be greater than 300 in ET and 3000 in NGO. The corresponding
SNR in the 33 mode is 90 and 900 for ET and NGO, respectively.
Even the least dominant 44 mode has SNRs of 20 and 200. 
The SNRs drop off as inverse distance and at redshifts $z\sim 1$-3,
from within which we can expect the event rate of massive black holes
to be in excess of several per year, the SNRs will be still large enough
that NGO should be able to carry out meaningful tests of GR for a 
large fraction of sources detected. Only very rare closeby events will
allow such tests in the case of ET.

\section{Tests of the no-hair theorem}\label{s:no-hair tests}
In this Section we will explore the ability of ET and NGO to test the 
no-hair theorem if they were to detect a ringdown signal from a 
black hole that resulted from the coalescence of a binary. We will begin
with a summary of the parameters used in our simulations followed by 
a description of the two different approaches that were used to test the
no-hair theorem. The first method uses consistency of the various
mode frequencies and damping constants with GR predictions and the
second method is a Bayesian model selection approach that addresses
which of a class of different models best describes the underlying
signal. We find that the latter approach is far more powerful in
testing the no-hair theorem.

The model selection approach followed in this paper is computationally
pretty expensive. We have therefore selected a small sample of signals
(two in ET and four in NGO) to assess how well future detectors are
able to test the no-hair theorem. The SNR study from the foregoing 
Section can be used to conclude how effective such tests are for a random
signal detected in our instruments.

\subsection{Choice of injection parameters}
\label{subsec:params}
The GW signal emitted from the QNM of a black hole ringdown as 
observed in a detector depends on the mass and spin 
of the final BH, the mass ratio of the progenitor binary 
and other extrinsic parameters that describe the 
orientation of the black hole, its distance from the
detector and its position on the sky. For the different 
sensitivity bands of ET and NGO we chose the following 
range of source parameters.
For ET, black holes of observed mass 500\,$\Msun$ and 1000\,$\Msun$ 
at luminosity distances, $D_{L}$, from 125\,Mpc and 225\,Mpc 
respectively, out to 6.63\,Gpc (corresponding to redshift 
$z \simeq 1$ \cite{Wright:2006}), were considered.  For NGO, black 
holes of observed mass $5 \times 10^{6} $ and $10^{8}\,\Msun$
at $D_{L}$ = 1 -- 59\,Gpc (with the upper limit corresponding 
to a redshift $z \simeq 6$), and black
holes of observed mass $10^{5} $ and $10^{6}\,\Msun$
at $D_{L}$ = 125\,Mpc -- 6.63\,Gpc, were considered.
In both cases, the source position ($\alpha,\delta$) was 
set to be directly above the detector at the time of observation. 

For all systems, the inclination and polarisation angles ($\iota,\psi$) were set to 
be ($\frac{\pi}{3}$,$\frac{\pi}{3}$) and the azimuth $\phi$ was taken to be zero. 
For the first four systems described, the mass ratio $q$ of 
the binary system prior to merger and the spin $j$ of the 
black hole after merger were set to $q=2$ and $j=0.61,$ respectively.
For the final two systems outlined, however, $q=10$ and $j=0.26$, to represent a more typical NGO binary system.
For the sake of using the phenomenological fit in table \ref{fitting_table} the binary components were 
assumed to be non-spinning.  It is important to note, 
especially for NGO, that the observed mass of the system 
is redshifted to be greater than the intrinsic mass of the 
system by a factor of $(1 + z).$ Here we report this 
observed, redshifted mass for injections and recovery.

For injections consistent with GR, parameters $\Delta\hat{\omega}_{lm} = 
\Delta\hat{\tau}_{lm} = 0$, whereas for non-GR signals, either $\Delta\hat{\omega}_{22}$ or 
$\Delta\hat{\tau}_{22}$ was varied in the range $-0.01 \ldots -0.1 $, 
with all other $\Delta\hat{\omega}_{lm} = \Delta\hat{\tau}_{lm} = 0$. 

\subsection{Constraining QNM parameters}
\label{subsec:GRsigs}
Having established the parameters of our test sources, we now move on to consider the two tests of GR.
The first method broadly follows the outline of Ref.\,\cite{Kamaretsos:2011} to estimate the parameters of each QNM in the ringdown signal. By choosing two of the measurements one can infer the true mass and spin of the black hole by excluding regions in the $M,j$ plane which are inconsistent with the measured parameters. The third measurement is then used to confirm the consistency of the inference. In the case of GR the allowed area will intersect both the previous areas as in Figure \ref{minset_GRconfidence}. Alternatively, if the signal is inconsistent with GR the intersections of the confidence regions will not agree, as in Figure \ref{minset_nonGRconsistency}.

In order to compute the constraints, we find the upper and lower 90\% probability interval of the marginal posterior probability distribution of the parameters $\Delta\hat{\omega}_{lm}$ and $\Delta\hat{\tau}_{lm}$.
In each case the injected values of these parameters were set to $0$, corresponding to the GR waveform. The priors for $\Delta\hat{\tau}_{lm}$ and $\Delta\hat{\omega}_{lm}$ were all uniform in the interval $[-1.0,\, 0.3]$ for all analyses, and the $M$, $j$ and $q$ parameters were set to their injected values. As we are estimating the mode parameters directly, through $\Delta\hat{\omega}_{lm}$ and $\Delta\hat{\tau}_{lm}$, the values of $M$ and $j$ parameters do not play any role in the results other than to set the centre of the prior range. The reason for the asymmetric choice of interval is that by allowing $\Delta\hat{\omega}_{22}$ to be greater than 0.3, the frequencies of the 22 and 33 modes collide, leading to strong correlations between the parameters of these modes upon recovery and confusion between the modes. Note that we allow the frequency of the 33 mode to vary downward into the band of the 22 mode, but this does not result in difficulties as the 22 mode is already within this band, making its identification easier. In all searches, the luminosity distance of the source and orientation angles are assumed known and fixed to the aforementioned values.

We chose the $\Delta\hat{\omega}_{22}$, $\Delta\hat{\tau}_{22}$ and $\Delta\hat{\omega}_{33}$ parameters to perform our consistency test as these are the three recovered parameters with the greatest precision for the signals we considered, and will therefore give the most stringent test.

We first injected signals using the GR waveform and performed parameter estimation on the three test parameters. The 90\% probability limits on $\omega_{lm}$ and $\tau_{lm}$ 
were projected in the ($M$,$j$)-plane to show visually the consistency test between the three modes, and that they agree at the injected value ($M$, $j$) within the measurement accuracy. Figure \ref{minset_GRconfidence} shows that for each system, the projections of $\omega_{lm}$ and $\tau_{lm}$ coincide at the same position, and the region of coincidence encloses the injected value of the mass and spin of the system, as expected.

To contrast, in Figure \ref{minset_nonGRconsistency} we show the corresponding plots 
where the injection is performed with deviations from GR of $\Delta\hat{\omega}_{22}$ of 
$-0.01$ and $-0.05$ for both ET and the $10^{6}\Msun$ NGO system, whilst for the $10^{8}\Msun$ NGO system, deviations of $-0.001$ and $-0.005$ were considered (all other parameters are taken to be the same).

This demonstrates the feasibility of the method for testing the no-hair theorem using consistency of the modes, when the signal is strong and the modes are clearly distinguishable. Of course, we may not be fortunate enough to observe BH ringdowns from such nearby sources, in which case we expect the power of the method to diminish.

\begin{figure*}[t]
\centering
$\begin{array}{cc}
\includegraphics[width=1.1\columnwidth]{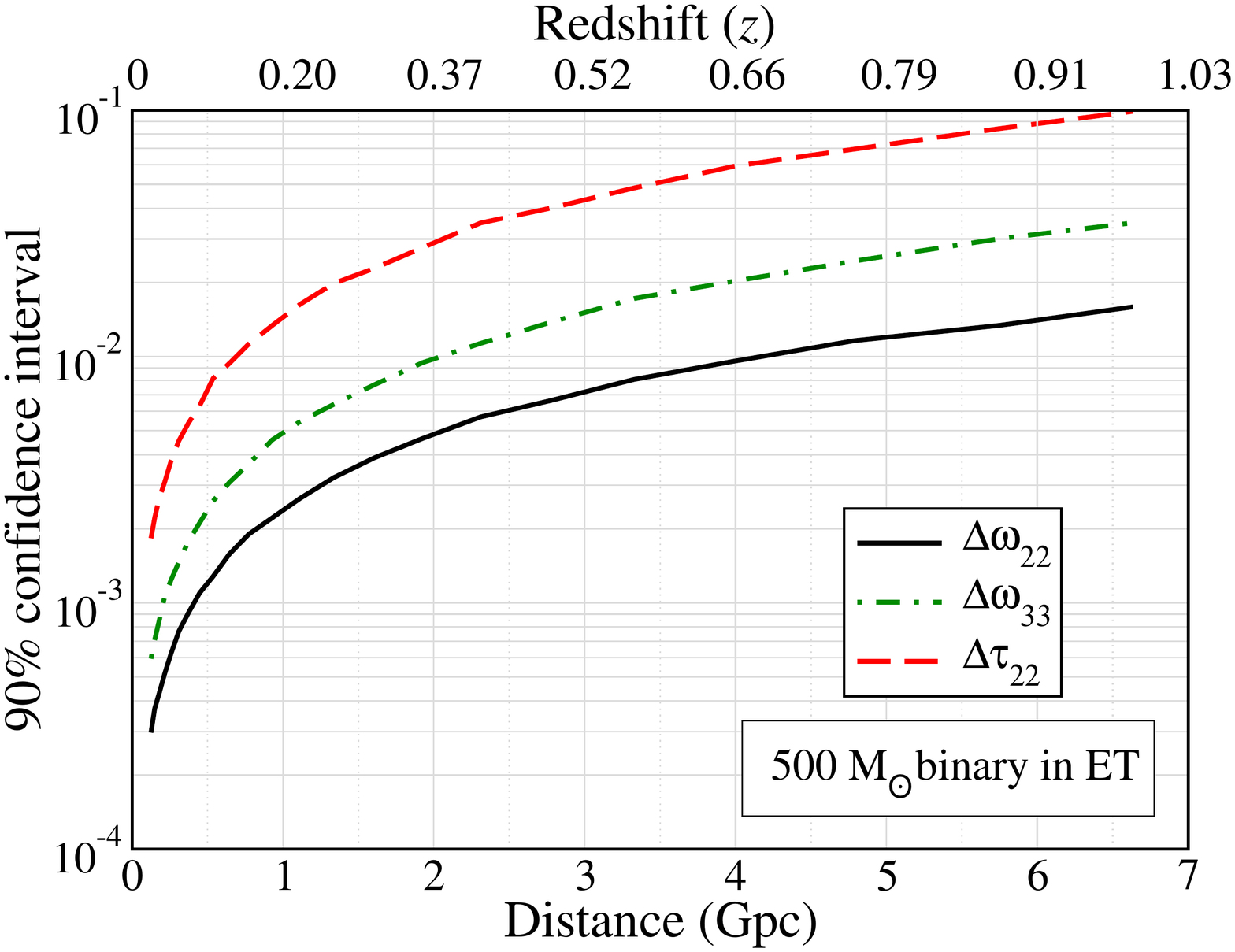}  & \hskip-0.9cm\includegraphics[width=1.1\columnwidth]{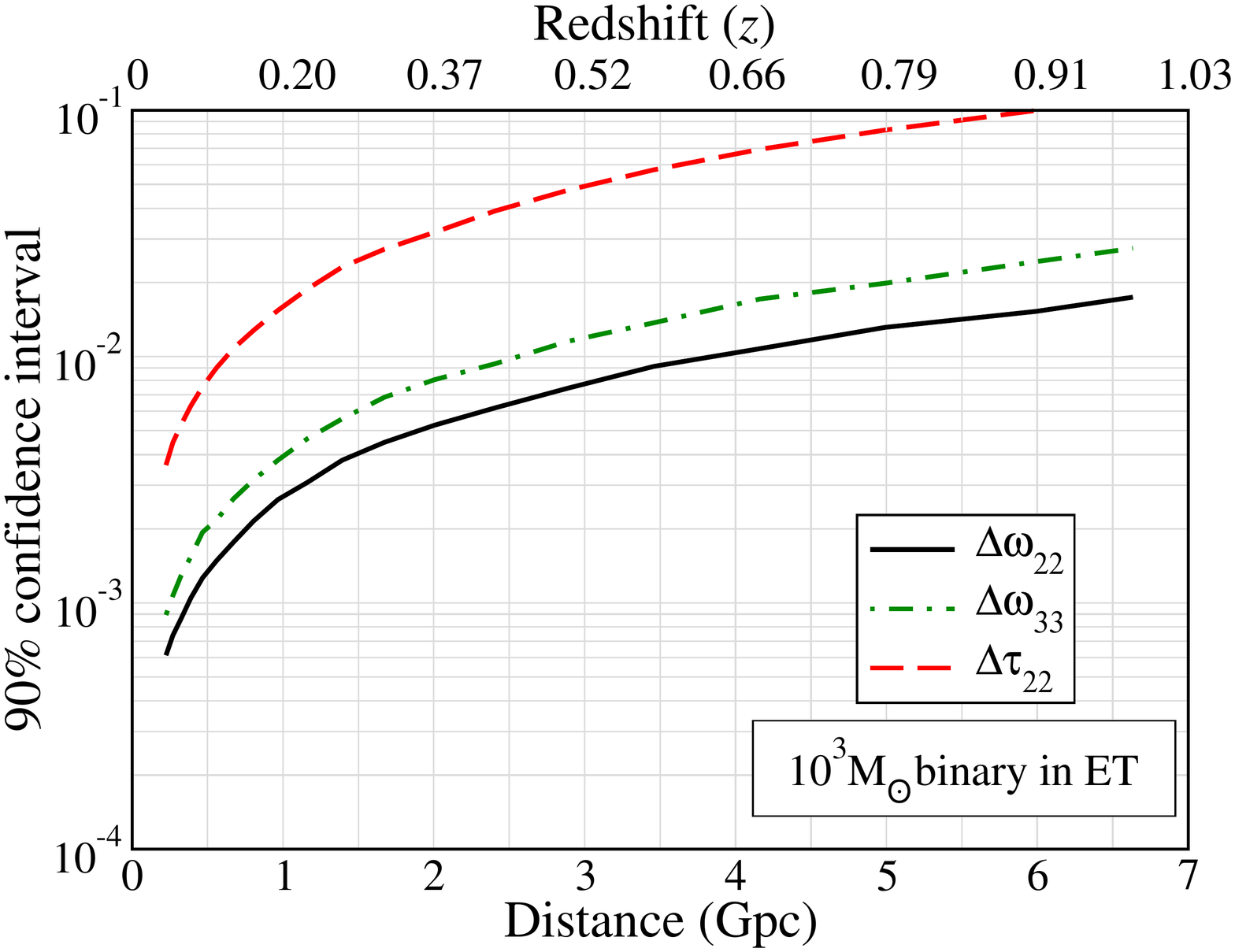} \\
\includegraphics[width=1.1\columnwidth]{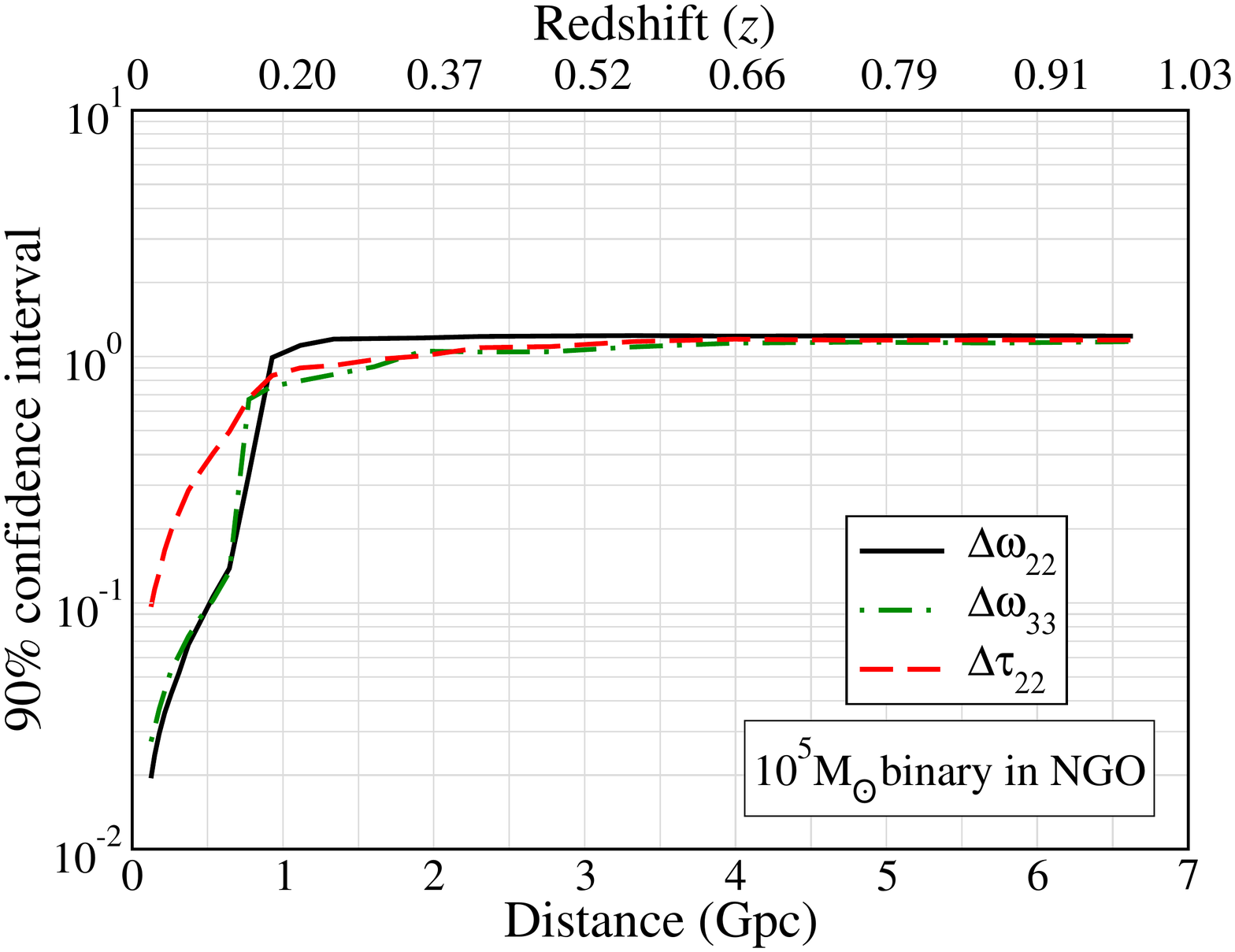} & \hskip-0.9cm\includegraphics[width=1.1\columnwidth]{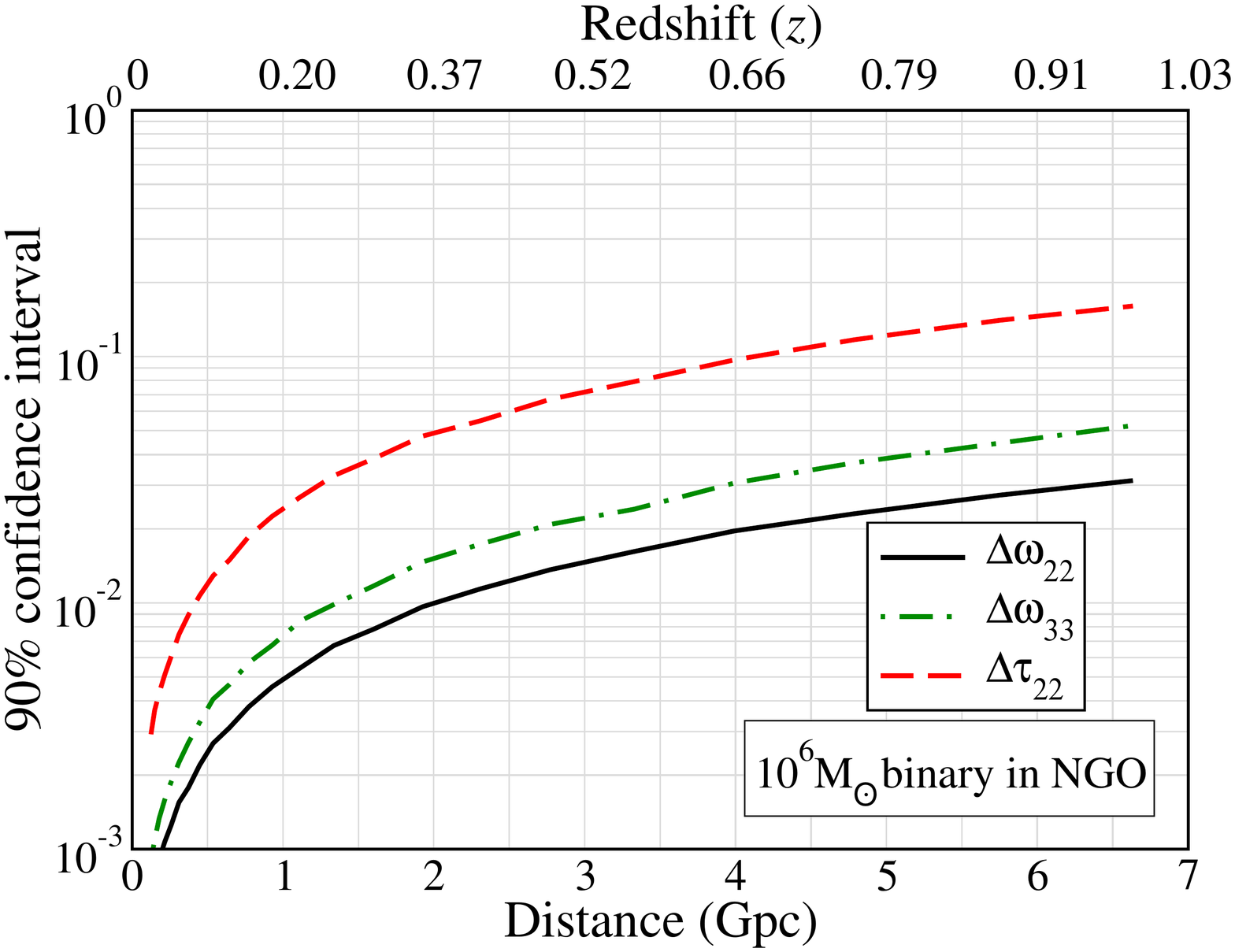} \\
\includegraphics[width=1.1\columnwidth]{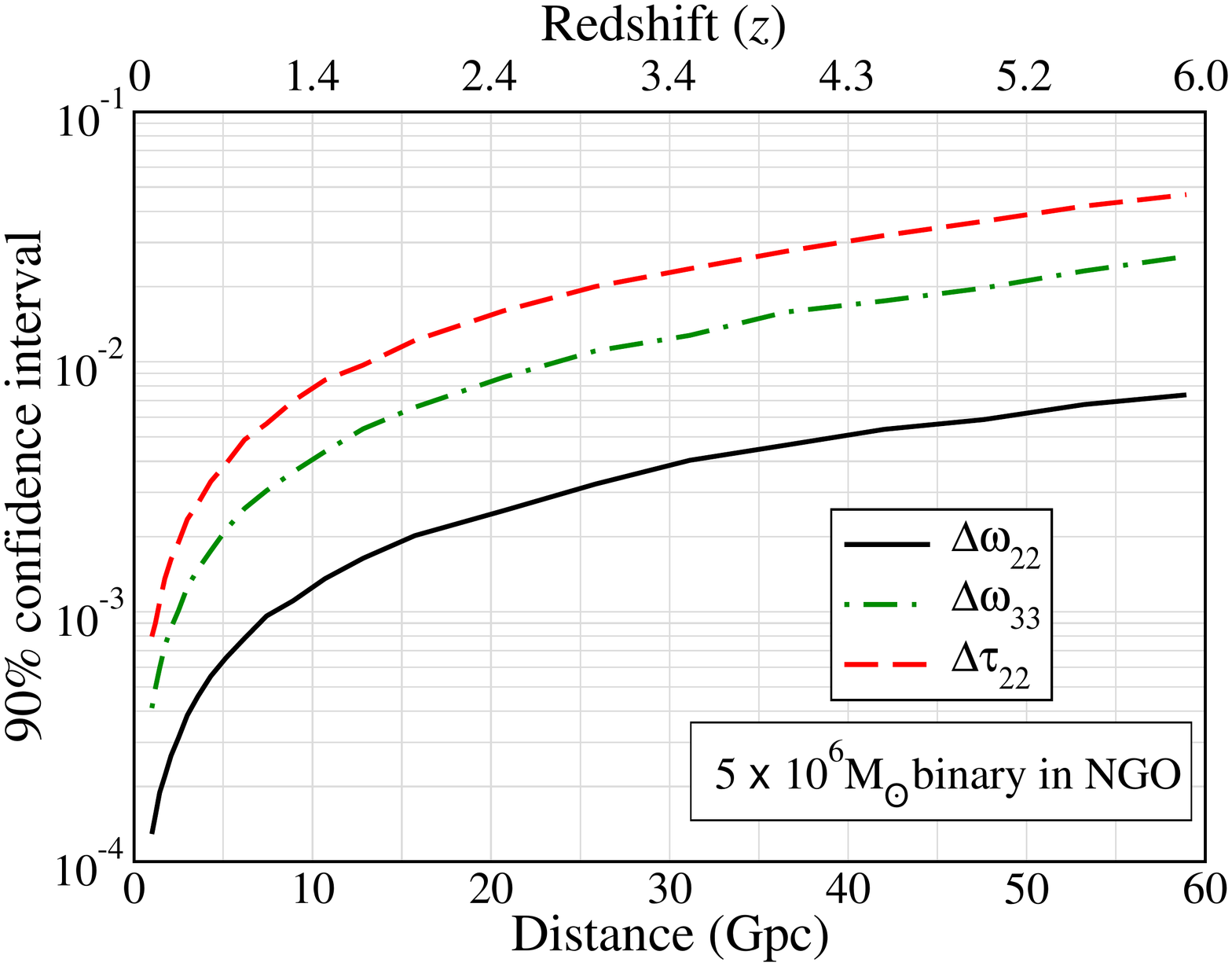} & \hskip-0.9cm\includegraphics[width=1.1\columnwidth]{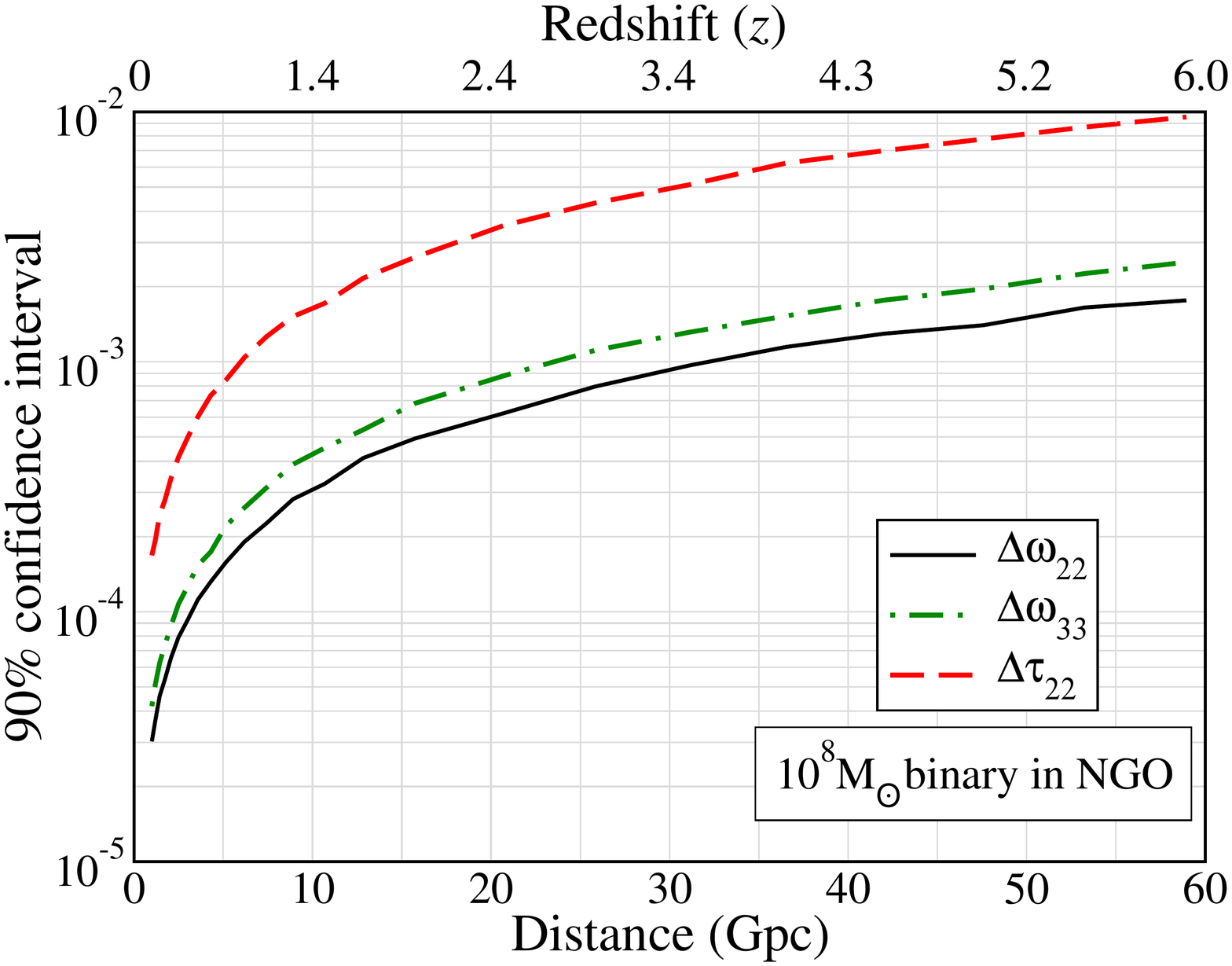} \\
\end{array}$
\caption{Width of the 90$\%$ confidence intervals for $\Delta\hat{\omega}_{22}$,$\Delta\hat{\omega}_{33}$ and $\Delta\hat{\tau}_{22}$ (blue, red and blue dotted lines respectively)
against luminosity distance for injections of $500$ (top-left), $1000$ (top-right), $10^{5}$ (middle-left), $10^{6}$ (middle-right), $5 \times 10^{6}$ (bottom-left) and $10^{8}\Msun$ (bottom-right) for ET (top) and NGO (middle and bottom) simulations. The limiting values in the $10^{5}\Msun$ plot are due to the posteriors expanding to fill the prior distributions when they are no longer resolvable.}
\label{minset_GRwidths}
\end{figure*}

To investigate the accuracy to which the mode parameters are resolvable we performed a set of injections at luminosity distances spanning the entire distance range quoted in Section \ref{subsec:params}, and estimated the $\Delta\hat{\omega}_{22}$, $\Delta\hat{\omega}_{33}$ and $\Delta\hat{\tau}_{22}$ parameters. For each black hole system considered, the width of the 90\% confidence intervals for the extracted values of $\Delta\hat{\omega}_{lm}$ and $\Delta\hat{\tau}_{lm}$ were plotted against luminosity distance in Figure \ref{minset_GRwidths}, for injections with GR waveforms. Our results are
in agreement with similar studies carried out in Ref.\,\cite{Kamaretsos:2011aa}.
From Fig.\,\ref{minset_GRwidths}, it can be seen that the width of the 90\% confidence intervals for the values of 
$\Delta\hat{\omega}_{22,33}$ and $\Delta\hat{\tau}_{22}$ increases with distance, as expected. $\Delta\hat{\omega}_{33}$ is extracted with considerably less accuracy than $\Delta\hat{\omega}_{22}$, because for a black hole system with $q = 2$, as considered here, the 33 mode is significantly less excited than the dominant 22 mode, and thus the 33 mode has a much lower SNR, resulting in poorer resolution.  By the same token, $\Delta\hat{\tau}_{22}$ is extracted with considerably less accuracy than both $\Delta\hat{\omega}_{22,33}$, as there is less extractable information from the mode decay times as opposed to the mode frequencies. In general, however, the relative weights and SNRs of each mode depends on the source parameters including orientation, and the noise curve of the detector.

We also see that ET can resolve $\Delta\hat{\omega}_{22}$ with errors of less than 1$\%$ at $D_{L} = 4$\,Gpc for the systems considered, in addition to resolving $\Delta\hat{\omega}_{22,33}$ and $\Delta\hat{\tau}_{22}$
with errors of 1.5\%, 2.5\% and $10\%$ for a black hole of 500\,$\Msun$ at $D_{L} = 6.63$\,Gpc.

NGO can resolve $\Delta\hat{\omega}_{22}$ to $\sim 0.6\%$ accuracy at $D_{L} = 59$\,Gpc ($z = 6$) for both $5\times10^6\,\Msun$ and $10^8\,\Msun$ systems, 
showing good results across the range of masses detectable.  $\Delta\hat{\tau}_{22}$ ($\Delta\hat{\omega}_{33}$) can be resolved to $\sim 3.5\%$ (1.5$\%$) and                  
$\sim 1\%$ (0.15\%) accuracy for $5\times10^6\,\Msun$ and $10^8\,\Msun$ systems respectively, at this distance.
This trend is mirrored for the $10^{6}\,\Msun$ system, with NGO able to resolve $\Delta\hat{\omega}_{22}$ to $\sim 2\%$ accuracy at $D_{L} = 6.63$\,Gpc ($z = 1$), and 
$\Delta\hat{\tau}_{22}$ ($\Delta\hat{\omega}_{33}$) to 10.5$\%$ (5$\%$) at this distance.  For the $10^{5}\,\Msun$ system however, resolution of $\Delta\hat{\omega}_{22}$ is poor even at
distances of $D_{L} = 1$\,Gpc, suggesting that it would not be possible to conduct these discriminatory tests using such systems, as they are not expected to exist at distances closer than
$1$\,Gpc.

The method outlined above to test the accuracy with which we can extract the three QNM parameters was applied to the case of non-GR signals ($\Delta\hat{\omega}_{lm}$ or $\Delta\hat{\tau}_{lm}$  set to some value within the range quoted in Section \ref{subsec:params}, with all other $\Delta\hat{\omega}_{lm} = \Delta\hat{\tau}_{lm} = 0$) from systems of $M = 500\,\Msun$ and $10^{8}\,\Msun$ in ET and NGO.
The QNM corresponding to these systems lie closest to the most sensitive frequency regions of the corresponding detectors, so this test is an optimistic case. 
Also, the distances used were from the lower end of the range quoted in Section \ref{subsec:params}.  
For each non-GR injection, the 90$\%$ probability limits on the values of
$\Delta\hat{\omega}_{22}$, $\Delta\hat{\omega}_{33}$ and $\Delta\hat{\tau}_{22}$ were extracted and used to project the corresponding confidence limits on 
$\omega_{lm}$ and $\tau_{lm}$ in the ($M, j$)-plane.

Figure \ref{minset_nonGRconsistency} shows that even for deviations from GR of as small as 1$\%$ (corresponding to $\Delta\hat{\omega}_{22} = -0.01$) for ET and the $10^{6}\,\Msun$ NGO system,
and as small as 0.1$\%$ (corresponding to $\Delta\hat{\omega}_{22} = -0.001$) for the $10^{8}\,\Msun$ NGO system, the projection of $\omega_{22}$
does not intersect consistently with the projections of $\tau_{22}$ and $\omega_{33}$, which were unchanged from their GR predicted values.  When an additional parameter dependence is introduced for one mode parameter, the mass and spin derived from this modified parameter will not be consistent with the 
mass and spin derived from the other unchanged parameters.  For a deviation of 5$\%$ for ET and the $10^{6}\,\Msun$ NGO system, and 0.5$\%$ for the $10^{8}\,\Msun$ NGO system, the projection 
of $\omega_{22}$ does not even touch the projections of $\tau_{22}$ and 
$\omega_{33}$, let alone intersect at the correct value.  This shows that in extracting the individual mode parameters from a QNM signal and projecting them in the 
$(M, j)$-plane, it is possible to determine if one, or many, of the mode parameters are not consistent with GR and have some parameter dependence other than on the mass and spin.

\subsection{Discriminating between models}
\label{models}
Although we have demonstrated the possibility of performing a consistency test 
between the QNM parameters by plotting them on the ($M$,$j$) plane, this method 
does not provide a quantitative measure of the consistency as such. In this 
section we seek to provide such a quantification by performing model selection 
on two competing models: the standard GR model for which we estimate $M$, $j$ 
and $q$; and the extended model where we estimate $M$, $j$ and $q$ but allow 
the individual mode parameters to vary away from their GR predicted values 
through the $\Delta\hat{\omega}_{lm}$ and $\Delta\hat{\tau}_{lm}$ parameters.
We shall see that even in the event where a source is too distant to allow the 
consistency method discussed in the previous section to be applied, it could still possible be
to perform model selection to distinguish between GR and non-GR models.

\begin{figure}[tbh]
\centering
\includegraphics[width=\columnwidth]{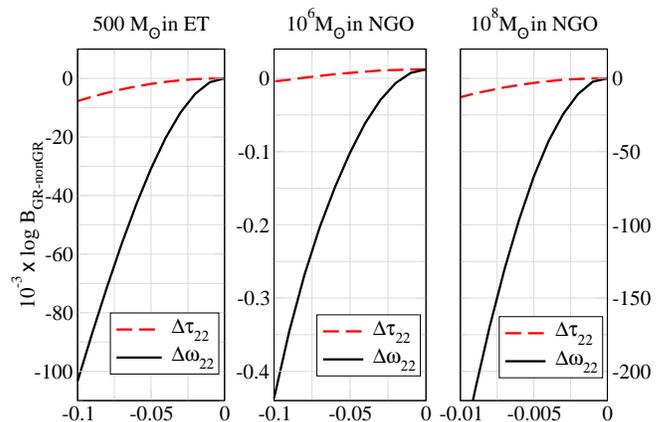}\\
\caption{The differences in evidence between the GR and non-GR models against $\Delta\hat{\omega}_{22}$ and $\Delta\hat{\tau}_{22}$ (blue and blue dotted lines respectively). (Top-Left) For injections of $M = 500$ at 125\,Mpc in ET, with SNR between $2\,664$ and $2\,888$. (Top-Right) For injections of $10^{8}\Msun$ in NGO at 1\,Gpc, with SNR between $114\,909$ and $115\,154$. (Bottom) For injections of $10^{6}\Msun$ in NGO at 125\,Mpc, with SNR between $1\,706$ and $1\,757$.}
\label{wtfixeddist}
\end{figure}

Bayesian model selection is a method of determining, for a given signal, whether the GR or non-GR
model is more likely by comparing the evidences for each. Note that unlike in the previous case where we estimated the QNM parameters directly, here we are estimating the physical parameters of the source along with deviations from the QNM parameters, where all QNM parameters are allowed to vary together.
The reason for this difference is that in estimating the QNM parameters directly we would not have a corresponding GR model with any free parameters unless we also searched over $M$, $j$ and $q$.
Because in the non-GR case we are measuring five parameters, including the best-determined ones, the resulting joint probability distributions are strongly correlated, although the presence of the $21$ mode and the assumption that $\Delta\tau_{33}=0$ help to break the degeneracy.

Here we compare evidences for both GR and non-GR signals injected into simulated ET and NGO data as described above, but calculate the evidence for each model using the nested sampling algorithm. The ratio of these evidences, known as the Bayes factor, quantifies the support for one model over the other provided by the data.
Full computation of the Bayesian evidence automatically takes into account the difference in dimensionality of the parameter space, which penalises the more complex non-GR model accordingly, as we shall see.

Figure \ref{wtfixeddist} displays how the Bayes factor between the GR and non-GR models changes as the deviation of the injected signal from GR increases. In the case of ET, we used the signal from a $500\,\Msun$ black hole at at distance of 125\,Mpc, giving a signal to noise ratio between $2\,644$ and $2\,888$.
The NGO simulation used both a $10^{8}\,\Msun$ system at a distance of 1\,Gpc, with signal to noise ratio between $114\,909$ and $115\,154$, and a $10^{6}\,\Msun$ system at a distance of 125\,Mpc, with signal to noise ratio between $1\,706$ and $1\,757$.
For ET and the $10^{6}\,\Msun$ NGO system, we varied both $\Delta\hat{\omega}_{22}$ and $\Delta\hat{\tau}_{22}$ in the range $-0.1$ to 0, whereas for the $10^{8}\,\Msun$ NGO system, we varied 
$\Delta\hat{\omega}_{22}$ and $\Delta\hat{\tau}_{22}$ in the range $-0.01$ to 0, and in each case computed the evidence for both GR and non-GR models.
As the deviation from GR increases, the evidence for the non-GR model increases whilst the evidence for the GR model decreases, causing the Bayes factor (the ratio of evidences for the two models in log space) to increasingly favour the non-GR model, as expected.  This is because as the deviation from GR increases, the data fits the GR model with less accuracy, but the parameter freedom of the non-GR model permits a consistency with the data. If we assume a threshold of $\log B_{GR,non-GR}=-10$ (corresponding to prior odds of $e^{10}$ in favour of the non-GR model), below which a deviation is considered significant, then with the source parameters above we are able to distinguish deviations at the level of 1\% with ET and NGO for the most sensitive $\Delta\hat{\omega}_{22}$ parameter.

\begin{figure*}[tbh]
 \centering
$\begin{array}{cc}
\includegraphics[width=1.1\columnwidth]{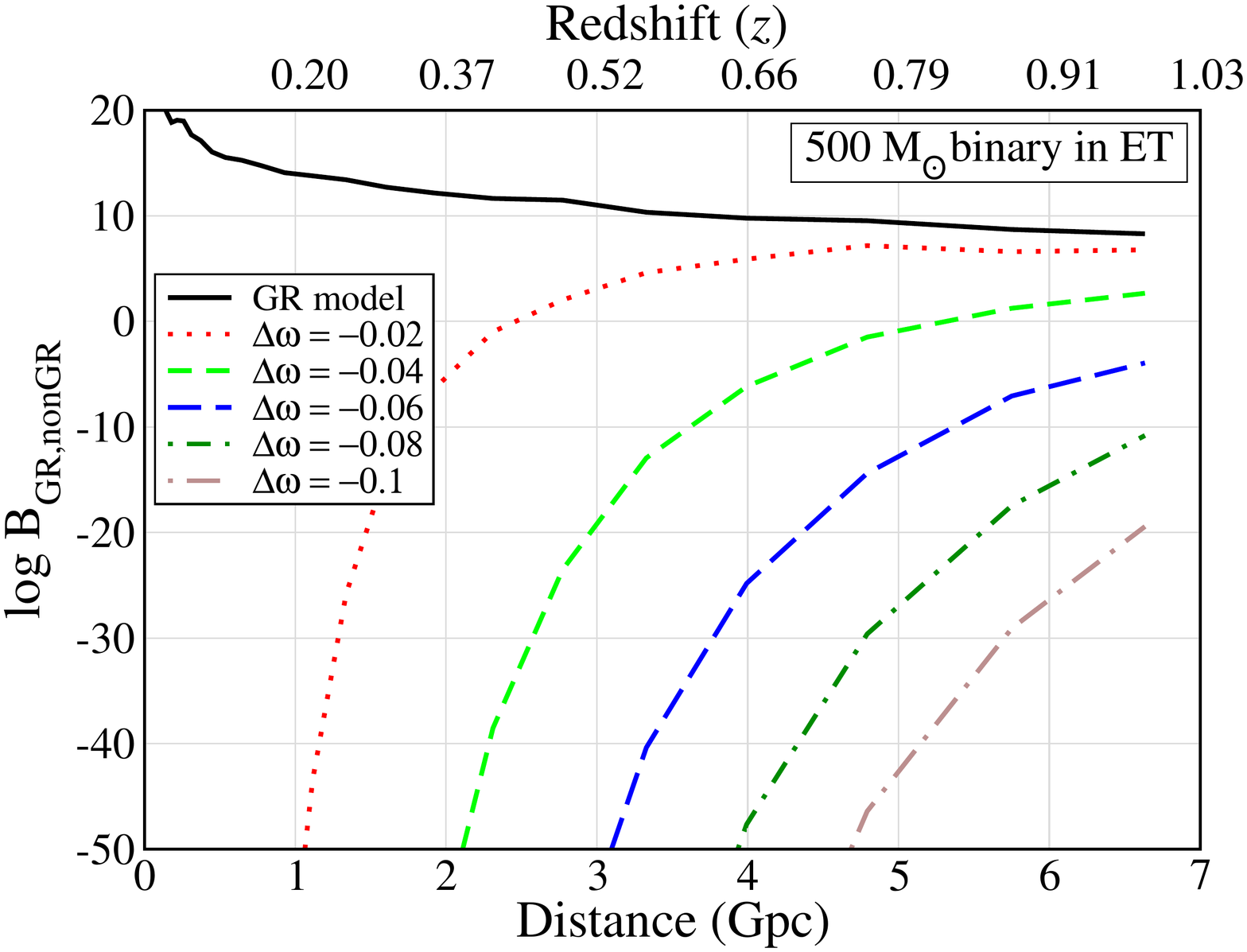}  & \hskip-0.9cm\includegraphics[width=1.1\columnwidth]{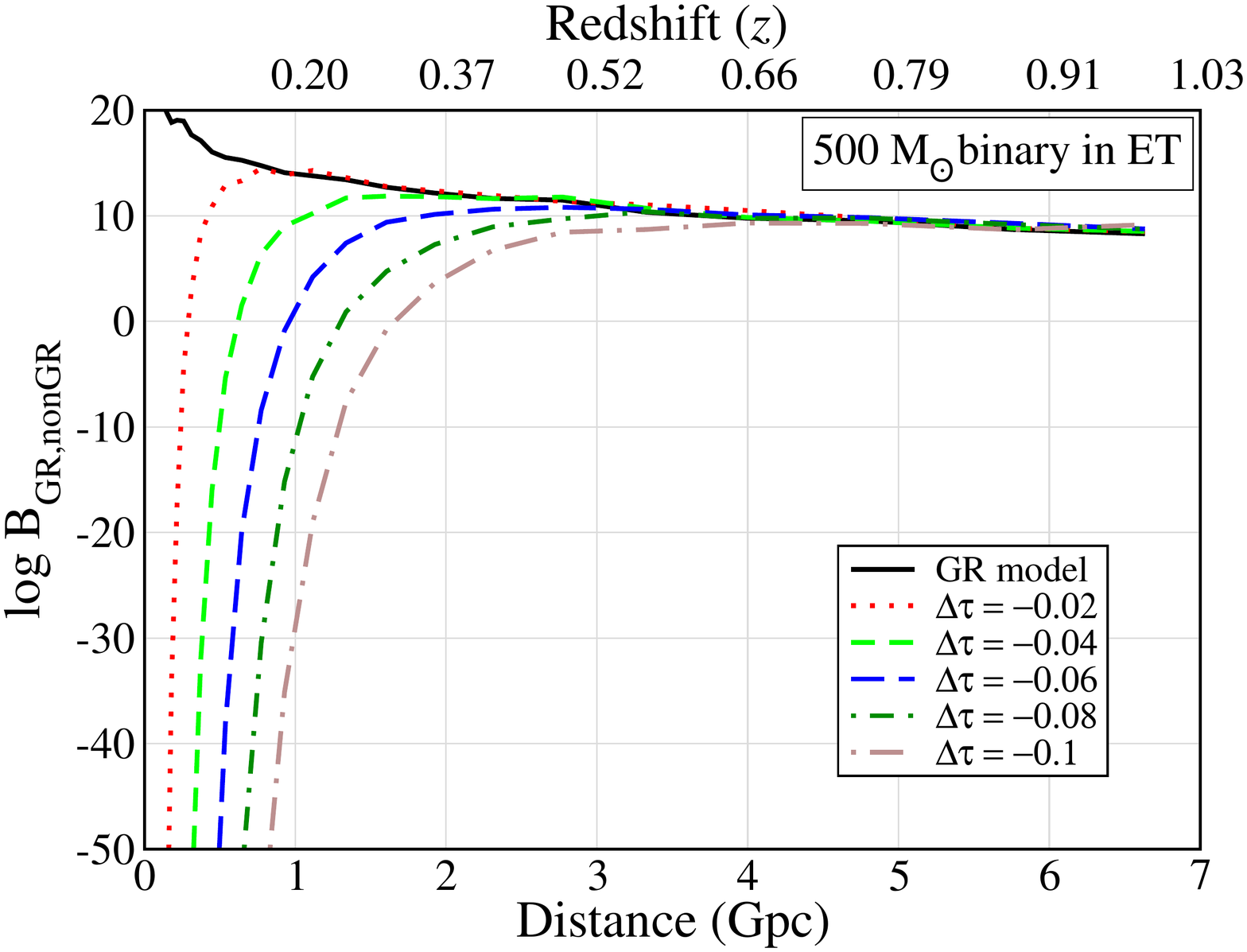} \\
\includegraphics[width=1.1\columnwidth]{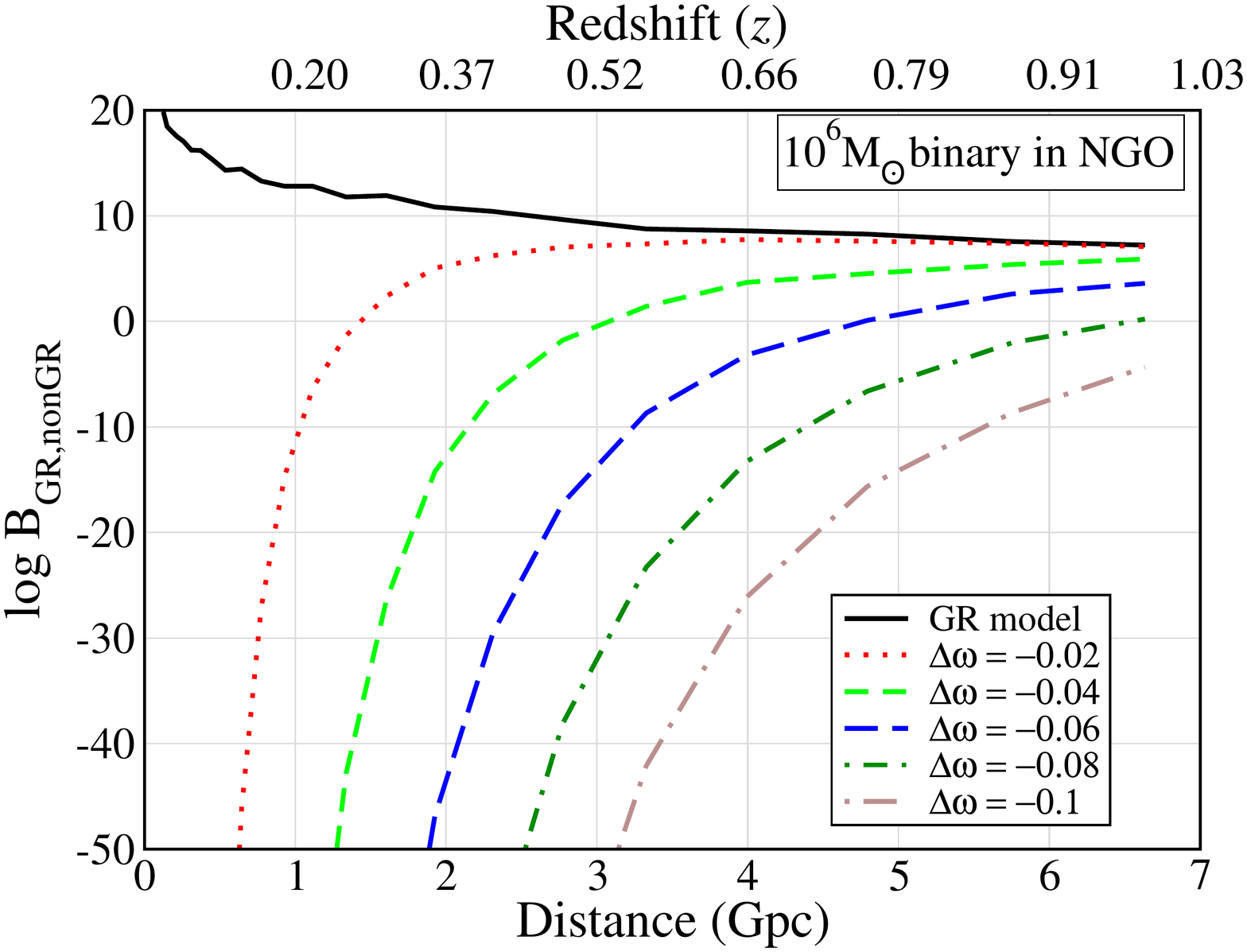} & \hskip-0.9cm\includegraphics[width=1.1\columnwidth]{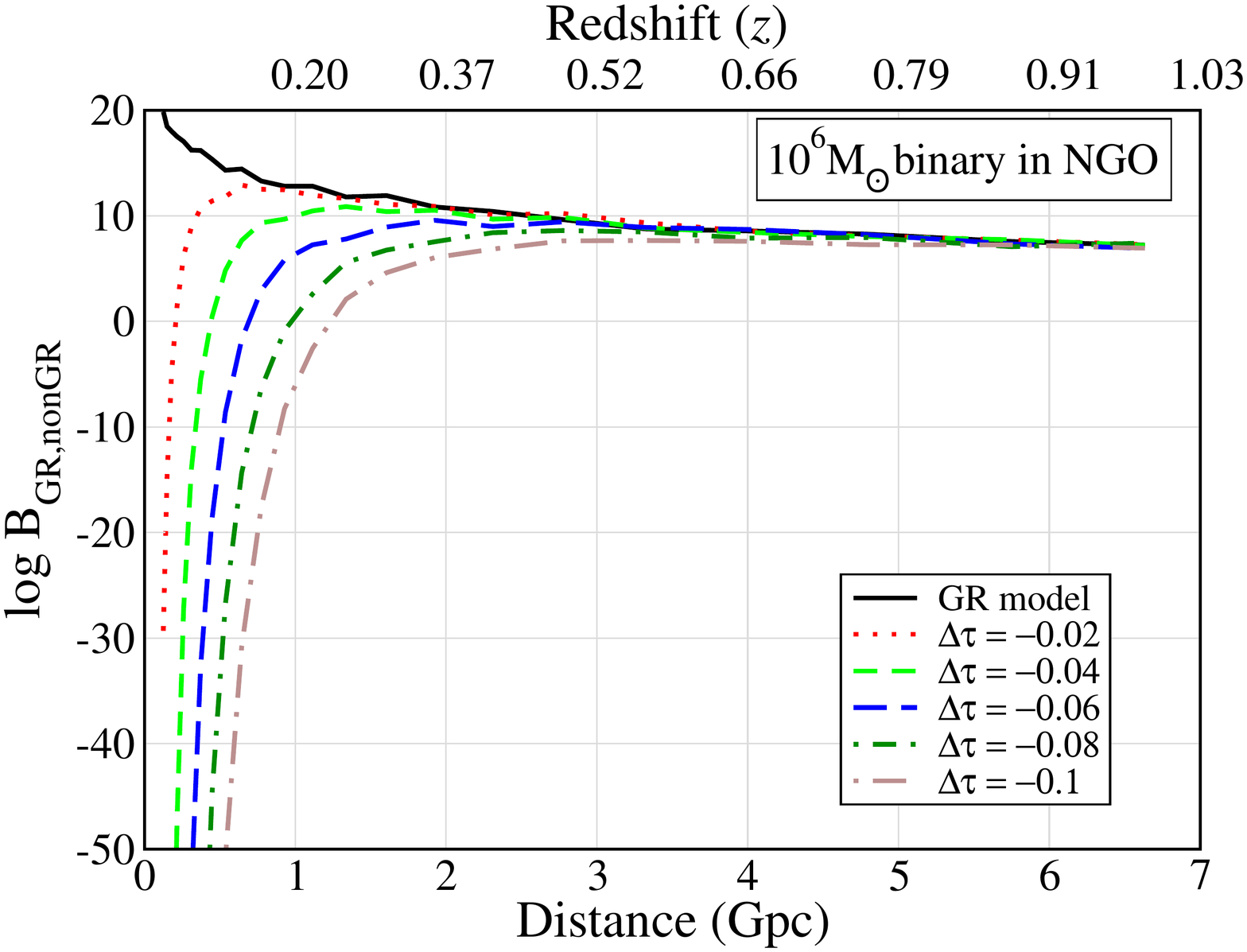} \\
\includegraphics[width=1.1\columnwidth]{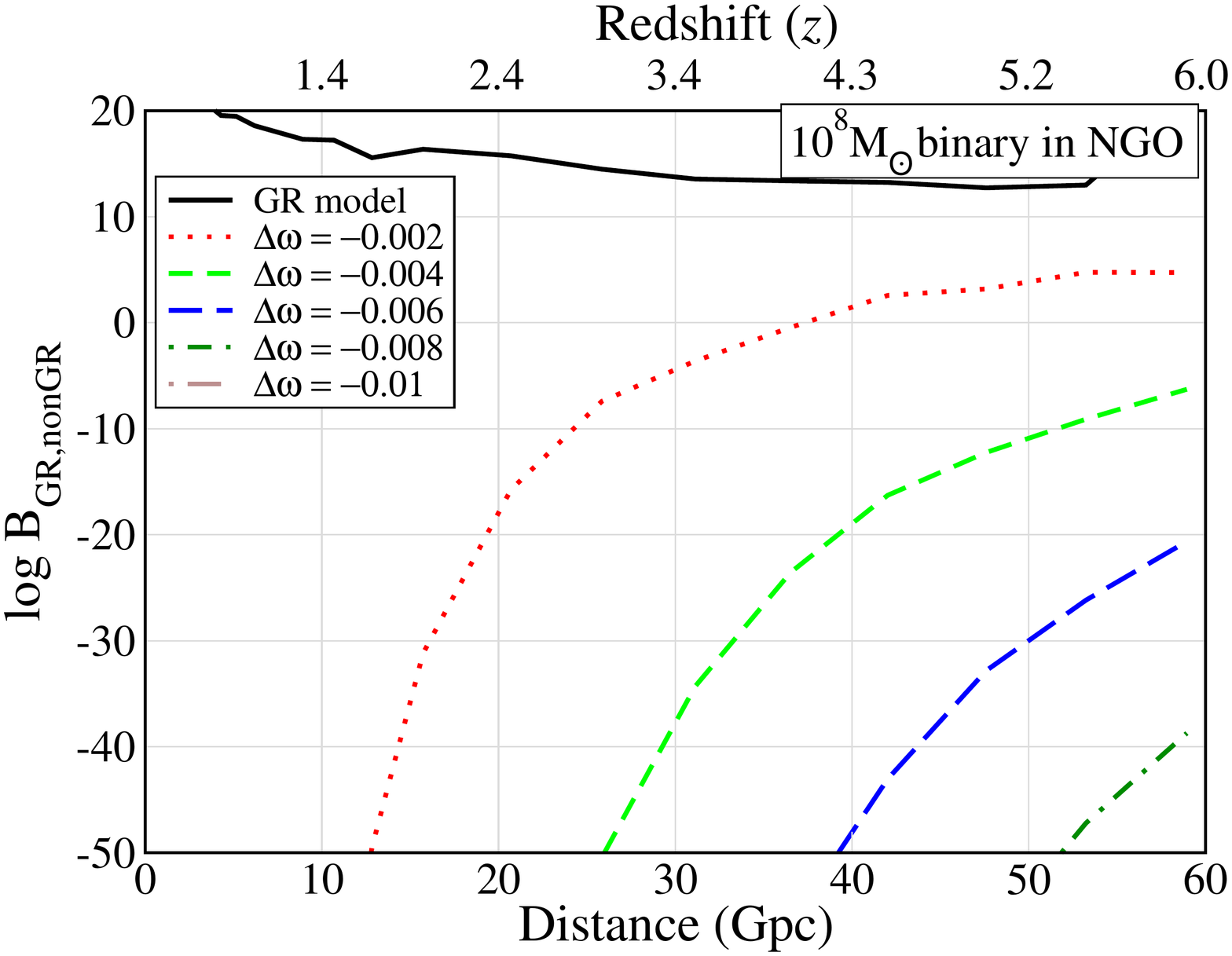} & \hskip-0.9cm\includegraphics[width=1.1\columnwidth]{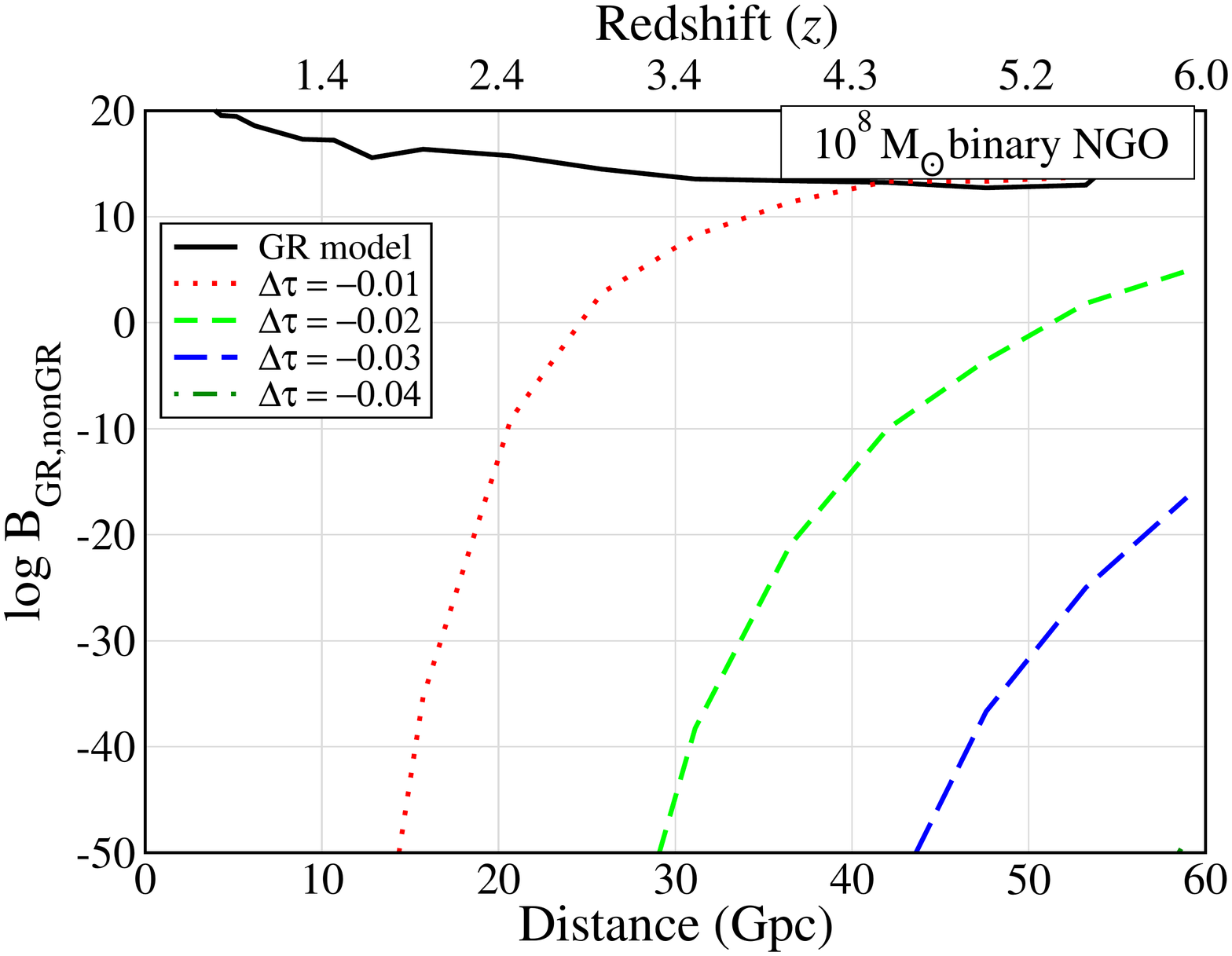} \\
\end{array}$
\caption{The differences in evidence for the GR and non-GR models is plotted against 
the luminosity distance to the source, for observation of ringdown signals from
a $500\,\Msun$ black hole in ET (top panels), $10^{8}$ and $10^{6}\,\Msun$ black hole
in NGO (middle and bottom panels repectively).  The non-GR model is simulated by varying either 
$\Delta\hat\omega_{22}$ (left panels) or $\Delta\hat\tau_{22}$ (right panels) as compared to
their GR values of 0.  $\Delta\hat \omega_{22}$ and $\Delta\hat \tau_{22}$ are varied over the 
range $0$ to $-0.1$ in steps of $-0.01$ in the case of ET and the $10^{6}\,\Msun$ NGO system, and over the range 
$0$ to $-0.01$ in steps of $-0.001$ in the case of the $10^{8}\,\Msun$ NGO system, while keeping all other
parameters as for GR. The different curves in each panel correspond to
different values of the non-GR parameter, starting with 0 for the top most curve
and changing by either $-0.01$ (top and bottom panels) or $-0.001$ (middle panels).
A difference in evidence of $-10$ or smaller is considered good enough to discriminate a non-GR model
from GR model.}
\label{bayesfactorsET}
\end{figure*}

As with the mode resolution of testing GR, the power of the model selection test decreases with increasing source luminosity distance since the signal visibility decreases.  The model
selection method was carried out for the sources above at distances spanning the entire range outlined in Section \ref{subsec:params}, and the Bayes factor between the GR and non-GR minimal set plotted as a function of distance for signals with different deviations from GR.  Figure \ref{bayesfactorsET} shows that for signals with 
larger deviations from GR, the maximal set non-GR model is favoured out to much greater distances.
We see also that for GR injections, the GR model is indeed preferred, since the simpler model fits the data adequately. When the distance is increased, the size of this effect decreases as the posterior distribution for the modified parameters gradually expands to fill the prior range. The power of the model selection method relies on the ability to exclude parts of the parameter space of the more complex model.

For the purposes of comparison, if we again impose a threshold on the Bayes factor of $\log B_{GR,non-GR}=-10$, we see that the non-GR model is favoured above the threshold at all distances for the $10^{8}\,\Msun$ system with at a deviation of only 0.6\%, whereas deviations as small as 0.1\% can be detected out to a distance of $\sim{}10.7$\,Gpc.  For the $10^{6}\,\Msun$ NGO system, the non-GR model is favoured above the threshold up to $\sim{}1$\,Gpc for a 2\% deviation and $\sim{}6$\,Gpc for a 10\% deviation.
For ET, we are able to cross the threshold up to $\sim{}1.61$\,Gpc for a 2\% deviation and 6.63\,Gpc for an 8\% deviation.

\section{Conclusions}\label{s:conclusions}
In this paper we have investigated how well quasi-normal modes could be
used to test general relativity. Our work is based on a quasi-normal mode
signal model \cite{Kamaretsos:2011} which assumes that the progenitor binaries 
are non-spinning. We have specifically looked at the ability of future ground-
and space-based detectors in testing the black hole no-hair theorem. More
specifically, we have investigated how well future interferometric gravitational
wave detectors can test deviations of quasi-normal mode frequencies and damping 
from their general relativistic values.

As expected, the $l=2$, $m=2$ mode of the signal is the most clearly measurable 
for the sources we considered, which can be used to infer the mass and spin 
of the black hole. Measurement of a third parameter, i.e. the frequency of the 
$l=3$, $m=3$ mode, is then used to confirm consistency between the modes. 
If the true signal is consistent with GR then the measurement of the different
parameters would be consistent with one another.  We find that a $10\%$ deviation 
in $\omega_{22}$ is clearly discernible for a $500\,\Msun$ source at 1.25\,Gpc in 
ET, as well as $10^{6}$ and $10^8\,\Msun$ sources at 1.25\,Gpc and 10\,Gpc, 
respectively, in NGO. Within the reach of NGO, the event rate could be pretty
high and so no-hair tests are promising in this case.
Binary coalescences of intermediate mass black holes within 1.25 Gpc
are highly unlikely and hence ET might not be able to carry out such tests,
except in the case of rare close by events.

In Section \ref{models}, therefore, we applied Bayesian model selection to obtain a 
more robust and quantitative measure of the consistency of the data with GR vs a 
generalised theory where the mode parameters depended on an extra parameter other
than the black hole mass and spin (i.e., `hairy' black holes).
Using this technique, we are able to measure deviations at the 10\% level 
in the $\hat{\omega}_{22}$ parameter out to $\simeq 6$\,Gpc for a 
$500\,\Msun$ source with the Einstein Telescope.  With NGO, we are able 
to measure deviations at the 10\% level at $6$\,Gpc with a $10^{6}\,\Msun$ 
source, and the 0.6\% level at $z \sim 5.1$ with a $10^8\,\Msun$ source. 

In the above analysis, we have assumed that the location and orientation of 
the GW source is known prior to the analysis---a significantly simplified 
analysis compared to the full problem of determining these parameters 
alongside the test of relativity.  Another limitation of our study is that
it is restricted a small number of systems with very specific source location,
orientation and mass ratio of the progenitor binary. Moreover, due to the lack
of accurate models, the study has used a very simple model of the ringdown 
signal that neglects the effect of initially spinning black holes of the 
progenitor binary.  The presence of spins could significantly alter the 
spectrum of the modes excited and their relative amplitudes, which could 
impact our ability to infer the mass ratio of the progenitor binary and 
the no-hair tests studied in this paper. Nevertheless, the methods we have 
developed in this work are equally applicable to the more general case, 
which should be performed as part of a follow-up study. Such a study will
be helpful in a more robust evaluation of the potential of future detectors
in testing the no-hair theorem.

\section{Acknowledgements}
The authors gratefully acknowledge support from the Science and Technology Facilities Council (STFC) UK, grant ST/H002006/1. We would like to thank Ioannis Kamaretsos for his input and advice on the simulations, Ilya Mandel for helpful comments, and the LIGO DataGrid clusters which allowed us to run the simulations used in this paper.

\bibliography{Bibliography}{}
\end{document}